\documentclass{aa}  

\usepackage{txfonts}

\usepackage[colorlinks=true, allcolors=blue]{hyperref}

\usepackage{graphicx}   
\usepackage[export]{adjustbox}

\newcommand{\revision}[1]{#1}

\def\gsim{\;\raise0.3ex\hbox{$>$\kern-0.75em\raise-1.1ex\hbox{$\sim$}}\;}

\def\lsim{\;\raise0.3ex\hbox{$<$\kern-0.75em\raise-1.1ex\hbox{$\sim$}}\;}
\def\gsim{\;\raise0.3ex\hbox{$>$\kern-0.75em\raise-1.1ex\hbox{$\sim$}}\;}
\begin{document}

\title{X-ray panorama of the SS~433/W50 complex by SRG/eROSITA}

  \author{
    Rashid~Sunyaev \inst{1,2}
    \and
    Ildar~Khabibullin \inst{3,2,1}
    \and
    Eugene~Churazov \inst{2,1}
    \and
    Marat~Gilfanov\inst{1,2}
    \and 
    Pavel~Medvedev\inst{1}
    \and
    Sergey~Sazonov\inst{1}
} 

\institute{
Space Research Institute (IKI), Profsoyuznaya 84/32, Moscow 117997, Russia
\and  
Max Planck Institute for Astrophysics, Karl-Schwarzschild-Str. 1, D-85741 Garching, Germany
\and
Universitäts-Sternwarte, Fakultät für Physik, Ludwig-Maximilians-Universität München, Scheinerstr.1, 81679 München, Germany
}

\abstract 
{    
~~~~
The Galactic microquasar SS~433 and the radio nebula W50 surrounding it present a prototypical example of a hyper-Eddington binary system shaping its ambient interstellar medium via energetic outflows.
In this paper, we present X-ray observations of the SS~433/W50 complex by the eROSITA telescope onboard the \textit{SRG} {space observatory}. These data provide images of the entire nebula characterized by a very large dynamic range and allow spectral analysis of the diffuse X-ray emission.
In particular, these data illustrate a close connection between the thermal and non-thermal components of W50 on scales ranging from sub-parsecs, represented by narrow X-ray bright filaments, to the entire extent of $\gtrsim 100\,{\rm pc}$ of the nebula. These data also allowed us to fully characterize a pair of nearly symmetric, sharp-edged, elongated structures aligned with the orbital axis of the binary system, which lack radio counterparts but are prominent in very-high-energy gamma-ray emission. The resulting multifaceted picture of the interaction between energetic outflows and the surrounding medium paves the way for future focused multiwavelength observations and {dedicated} numerical simulations.}

\titlerunning{SS433/W50 with SRG/eROSITA}
\maketitle

\section{Introduction}
\label{sec:intro}
~~~~SS~433 is a unique close stellar binary system in our Galaxy (located at $l,b=39.7^\circ,-2.2^\circ$ and $d_{\rm SS433}\approx 5$ kpc) composed of a relativistic compact object (a black hole or a neutron star), which sustainably accretes matter from its companion at a hyper-Eddington rate {\citep[see ][ for reviews]{2004ASPRv..12....1F,2025PhyU...68.1042C}}. Having first been listed in the catalog of point-like H$\alpha$ emitters \citep{1977ApJS...33..459S}, it quickly became one of the most studied astrophysical objects thanks to its very peculiar optical spectrum \citep{1979ApJ...230L..41M} featuring pairs of narrow (with a half-opening angle of $\theta_{\rm j}\lesssim2^\circ$) {blue- and redshifted} lines (with respect to their rest-frame positions), with the shifts varying periodically in time \citep[for a review of early observational results see][
]{1984ARA&A..22..507M}. A kinematic model associates this emission with a pair of narrow relativistic jets with a constant bulk velocity of a quarter of the speed of light ($\beta_{\rm j}=0.26$c) that change direction in a regular precession manner with a period of $P_{\rm prec}\approx 162$ days \citep{1979Natur.279..701A,1979MNRAS.187P..13F}. This period is much longer than the period of orbital modulation {($P_{\rm orb}=13.08$ days)} and is interpreted as being due to the precession of a thick accretion disk with an amplitude of $\Theta_{\rm prec}\approx$21$^\circ$ and mean axis inclined by the angle $i=78^\circ$ to the line of sight \citep[e.g., ][]{1980ApJ...236L.127K,1980ApJ...238L.129S,1980MNRAS.193..707W}. Subsequent detection of extended and variable radio emission \citep{1979Natur.282..483S} perfectly described by the precessing jets pattern fully confirmed the picture of twin relativistic outflows \citep[][]{1981ApJ...246L.141H,1981Natur.290..100H,2004ApJ...616L.159B}. Moreover, the proper motion of the individual radio-emitting blobs allowed one to constrain the distance to the source with a remarkable 10\% accuracy; i.e., $d_{\rm SS~433}=5\pm0.5$ kpc \citep{1981ApJ...246L.141H, 2004ApJ...616L.159B,2013ApJ...775...75M}. 

Despite indications of both quasi-regular \citep[nutation, ][]{1982ApJ...260..780K} and sporadic \citep[jittering, ][]{2010PASJ...62..323K} deviations from the predictions of this simple model as well as the presence of (relatively short) periods of time when the Doppler shifts of optical emission lines change significantly \citep[e.g., ][]{2011MNRAS.417.2401B,2022AstL...48..389M}, all the derived parameters of the model stay remarkably stable over more than 40 years of continuous monitoring observations \citep{2018ARep...62..747C}. This stability likely reflects extreme robustness of the mass transfer occurring in the {hyper-Eddington} regime \citep{1981VA.....25...95V,2017MNRAS.471.4256V} with the estimated accretion rate, $\dot{M}\gtrsim 10^{-4}M_{\odot}~{\rm yr}^{-1}\sim 1000 \dot{M}_{\rm Edd} (M_{\rm X}/3M_{\odot})$ \citep[e.g., ][]{1981SvA....25..315S,1981MNRAS.194..761C,1984MNRAS.210..279L,2006A&A...445.1041F}, where $M_{\rm X}$ is the mass of the compact object, $M_{\odot}$ is the solar mass, and $\dot{M}_{\rm Edd}$ is the critical {Eddington} mass-accretion rate. {The compact object is most likely a black hole with $M_{\rm X}\gtrsim8M_{\odot}$ \citep[][]{2023NewA..10302060C}, although the possibility of it being a neutron star ($M_{\rm X}<3M_{\odot}$) has also been discussed \citep[e.g.,][]{2011PZ.....31....5G,2018AstL...44..390M}. }

 Since X-ray emission coming from the innermost parts of the {hyper-Eddington} accretion disk is expected to be strongly beamed in the axial direction \citep[][]{1973A&A....24..337S,2006MNRAS.370..399B,2007MNRAS.377.1187P}, we are not able to observe it directly, given the high inclination angle of the system. Although X-ray emission from this source was detected early on \citep{1976MNRAS.175P..39S,1978ApJS...38..357F}, \revision{ it was also found \citep[e.g.,][]{1984ApJ...277..286G} that its apparent luminosity, $\sim10^{36}~{\rm erg~s^{-1}}$, is very low compared to the Eddington luminosity, $L_{\rm Edd}\approx10^{39}$~erg~s$^{-1}$, of an $M_{\rm X}=8M_{\odot}$ compact object}. As a result, SS~433 has long been considered as a beamed away prototype of ultra-luminous X-ray sources \citep[ULXs; for a review see][
 ]{2017ARA&A..55..303K} observed in normal star-forming galaxies \citep{2001IAUS..205..268F,2006MNRAS.370..399B,2007MNRAS.377.1187P}, although possible indications of the "hidden" central emission are 
 yet to be confirmed \citep{2010MNRAS.402..479M,2016MNRAS.457.3963K,2019AstL...45..282K,2021MNRAS.506.1045M,2023A&A...669A.149F}.
Instead, what dominates the X-ray emission of the system is again thermal emission from a twin pair of the baryonic jets having almost identical orientation and bulk velocity to the optically emitting jets \citep{1986MNRAS.222..261W}. Numerous pairs of blue- and redshifted narrow emission lines of highly ionized heavy elements (primarily silicon, sulfur, iron, and nickel) allow one to build a multi-temperature model of the ballistically expanding and cooling flow \citep{1986MNRAS.222..261W,1996PASJ...48..619K,2002ApJ...564..941M,2005A&A...431..575B,2010MNRAS.402..479M,2013ApJ...775...75M,2016MNRAS.455.1414K}, which first becomes visible when its temperature is above 20 kiloelectronvolt (keV), then cools down below 1 keV, and likely fragments later on due to thermal instability \citep{1988A&A...196..313B}. Detailed spectroscopic modeling of this emission \citep[e.g.,][]{2016MNRAS.455.1414K,2019AstL...45..299M} in combination with the eclipsing observations of the jets by the companion star \citep{1987MNRAS.228..293S,1989PASJ...41..491K,2006A&A...460..125F,2006ApJ...650..338L,2016AstL...42..517A} enable very robust determination of the jets' physical properties (their so-called "biometry", consisting of their size, opening angle, velocity, temperature, and the mass flux). Thanks to this we know the kinetic luminosity of these outflows with certainty; it exceeds $10^{39}~{\rm erg~s^{-1}}$ \citep[e.g.,][]{2019AstL...45..299M}, confirming the necessity of a {hyper-Eddington} "engine" releasing large amount of energy in the form of relativistic outflows, consistently with the early theoretical \citep{1973A&A....24..337S} and modern numerical \citep[e.g.,][]{2011ApJ...736....2O,2019ApJ...880...67J,2022PASJ...74.1378Y,2025MNRAS.540.2820F} predictions.
 
 In contrast to the radiation freely escaping the system, the kinetic energy of the launched outflows cannot be easily transported away from the source and has to be deposited into the surrounding interstellar medium. The estimated total amount of this energy {exceeds $10^{51}$ erg for a lifetime of the system at the level of $\gtrsim$30,000 yr} \citep[e.g.,][]{1980A&A....81L...7V}. This energy is at least on par with the energy of the initial supernova explosion that accompanied the birth of the compact object, meaning that quite a large region of $\gtrsim 10$ pc in its vicinity can be shaped by this later activity \citep[e.g.,][]{1980ApJ...238..722B,1980MNRAS.192..731Z,1981VA.....25...95V,2000A&A...362..780V,2008MNRAS.387..839Z,2011MNRAS.414.2838G, 2020MNRAS.495L..51C, 2024A&A...688A...4C}. 
 
 Indeed, SS~433 is known to be surrounded by the famous giant ($\sim200\times100$ pc) radio nebula W50 \citep{1958BAN....14..215W,1969MNRAS.143..407H,1980A&A....84..237G, 1987AJ.....94.1633E,1998AJ....116.1842D, 2011MNRAS.414.2828G,2018MNRAS.475.5360B,2021PASJ...73..530S}; the latter has a suggestive, elongated morphology, with the elongation axis being co-aligned with the symmetry axis of the jets' precession cone. The asymmetry of the lobes, or ``ears'', of the nebula can be readily explained by the noticeable stratification of the surrounding medium, given that the system is located at a height of $h_{\rm SS~433}=d_{\rm SS~433}\sin(b)\approx200$ pc below the Galactic plane and has the size comparable to the scale height of the exponential cold gas disk \citep[e.g.,][]{2000A&A...362..780V,2008MNRAS.387..839Z,2011MNRAS.414.2838G}. 
The interior of the nebula is also known to shine in soft thermal X-rays \citep{1996A&A...312..306B,2007A&A...463..611B,1997ApJ...483..868S,2024ApJ...975L..28C}, hard nonthermal X-rays
 \citep[][]{1980Natur.287..806S,1983ApJ...273..688W,1994PASJ...46L.109Y,1999ApJ...512..784S,2000AdSpR..25..709N,2022ApJ...935..163S,2024ApJ...961L..12K}, and filamentary H$\alpha$ emission \citep{1980ApJ...242L..77K,1980PASP...92..259S,2007MNRAS.381..308B,2010AN....331..412A,2017MNRAS.467.4777F,2021MNRAS.506.4263R}, which is indicative of the high-velocity shocks propagating through a relatively dense interstellar medium.  

The nature of the extended X-ray jets (EXJs) remains more puzzling, although their morphology, spectra, and polarization all point to the synchrotron origin of the emission \citep[e.g.,][]{2022ApJ...935..163S,2024ApJ...961L..12K}. Most recently, very-high-energy (VHE) emission from these structures has been detected and mapped \citep[e.g.,][]{2018Natur.562...82A,2024ApJ...976...30A,2024Sci...383..402H,2024arXiv241008988L}, demonstrating that ultra-relativistic electrons with energy above 100 teraelectronvolt (TeV) are accelerated in the system \citep[e.g.,][]{2020ApJ...904..188K}, probably as a consequence of internal shocks and recollimation in the energetic trans-relativistic outflows \citep{2025PhRvD.112f3017B}. The overall production of cosmic rays by such systems ({super-Eddington} accretors with powerful outflows) might be of importance for the total CR budget and local variations in its spectrum from place to place \citep[e.g.,][]{2002A&A...390..751H}, particularly in the petaelectronvolt (PeV)
regime \citep[e.g., ][]{2025A&A...698A.188P,2025ApJ...989L..25W,2025PhRvD.112f3017B}.

Although the paradigm of ``a {super-Eddington} X-ray binary powering a nebula with energetic outflows'' appears to be well established and consistent with the most salient observational properties of the SS~433/W50 complex, the details of almost all aspects of this interaction remain unclear. Indeed, all the phenomena directly associated with the central source can only be traced up to a distance of 0.1 pc from it, where the last signatures of the corkscrew pattern of the precessing jets are observed in radio and X-rays \citep[][]{2005MNRAS.358..860M,2008ApJ...682.1141M,2017AstL...43..388K,2025PASJ...77.1113S}. A model invoking steady deceleration and disruption of the jets can, in principle, account for such behavior \citep[][]{2014A&A...562A.130P,2017A&A...599A..77P}, but then the reappearance of the ``jets'' tens of parsec away from the central source with velocities similar to those at the injection
might be problematic. Morphology of the soft X-ray emission, best mapped by the \textit{ROSAT} mosaic observations \citep{1996A&A...312..306B}, is also quite puzzling, with large global brightness variations and thin filamentary structures, which are not commonly observed in similar situations of supernova remnants (SNRs, which typically feature shell-, plerion- or mixed types of morphology). Deep pointed observations with \textit{XMM-Newton}, \textit{Chandra}, \textit{Suzaku}, and \textit{NuSTAR} were capable of covering certain parts of the nebula \citep[e.g.,][]{2005A&A...431..575B,2022ApJ...935..163S,2024ApJ...975L..28C,2025ApJ...993L..24T}, but a uniform and sensitive coverage of the entire nebula with adequate spectroscopic capabilities is missing.

Here, we present results of X-ray observations of the SS~433/W50 complex by the eROSITA telescope \citep{2021A&A...647A...1P} on board the \textit{SRG} {space observatory} \citep{2021A&A...656A.132S}, which were performed during its performance-verification (PV) phase. The obtained mosaic provides us with X-ray images of the entire extent of W50 with unprecedented dynamic range, bringing X-ray data on par with the radio and other multiwavelength data, and, in addition, delivers exquisite spectral information for the rich variety of diffuse structures. These data demonstrate a close connection between the thermal and nonthermal components of W50 on scales ranging from subparsec (thickness of narrow X-ray and radio bright filaments), to the entire $\gtrsim100$~pc extent of the nebula. These data also enable the characterization of the pair of nearly symmetric, sharp-edged, elongated structures aligned with the orbital axis of the binary system ("extended X-ray jets"). 

The structure of the paper is as follows. We describe the observations and the obtained data in Sect.~\ref{s:data}; results of the broad-band and spectrally resolved imaging are presented in Sect.~\ref{s:broadband} and Sect.~\ref{s:specresolved}, respectively; a spectroscopic analysis of the emission from several selected regions follows in Sect.~\ref{s:spectroscopy}; and the obtained global picture is discussed in Sect.~\ref{s:discussion} and summarized in Sect.~\ref{s:conclusions}.

\section{Data}
\label{s:data}

\begin{figure*}
    \centering
    \includegraphics[width=0.522\linewidth,trim=2cm 6.5cm 4.5cm 5.5cm,clip]{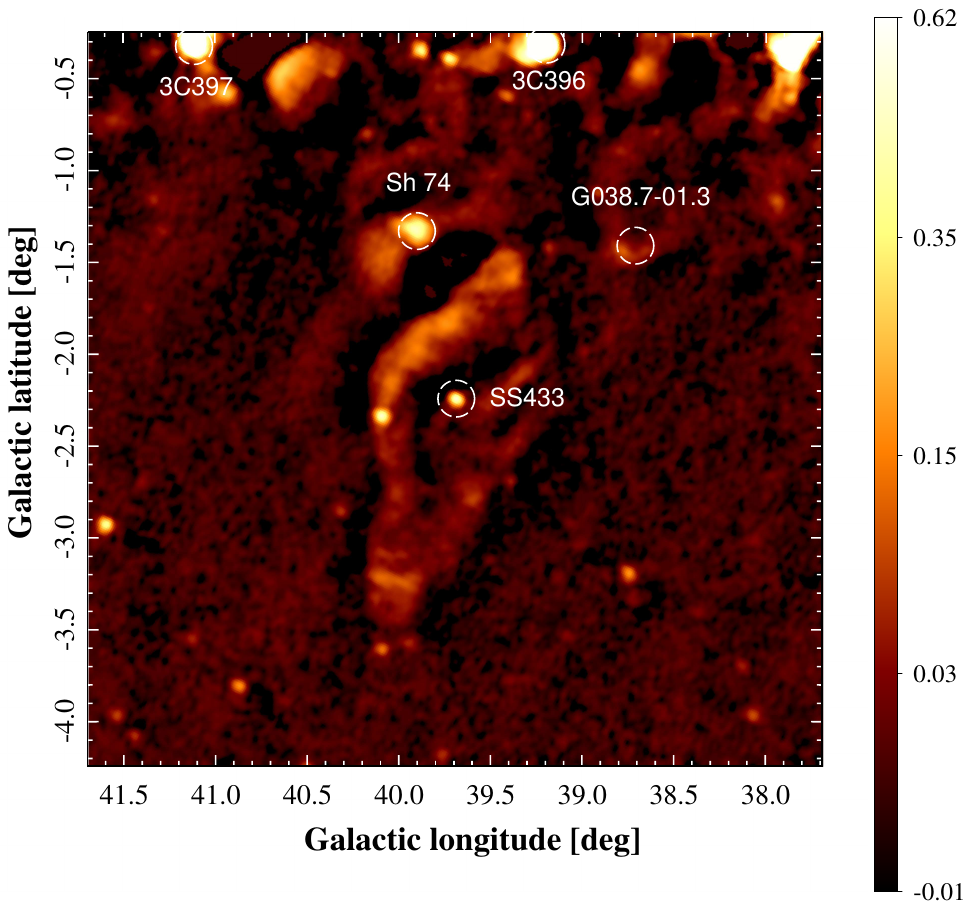}
        \includegraphics[width=0.47\linewidth,trim=3.5cm 6.5cm 4.5cm 5.5cm,clip]{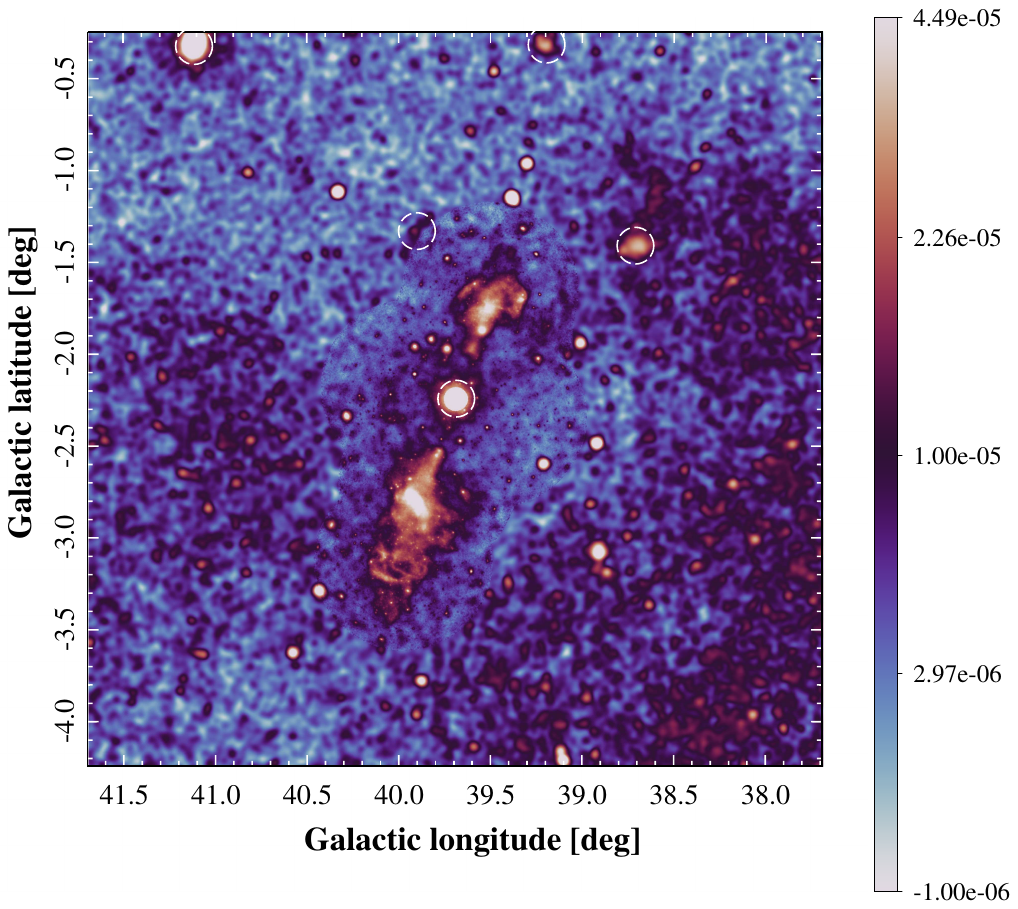}
    \caption{Global radio and X-ray views of $4\times4$ degree region centered on SS~433 {(both images are in Galactic coordinates)}. 
    {\bf Left:} Radio image {(square-root scale)} by Green Bank telescope at 6cm (4.85 GHz, \citealt{1996ApJS..103..427G}). {\bf Right: } \textit{SRG}/eROSITA (particle background-subtracted and exposure-corrected) 0.3-2.3 keV X-ray image {(log scale, two-orders-of-magnitude range)} of the W50 field, which combines the PV and all-sky data accumulated over four consecutive scans after 
    smoothing using a Gaussian beam with $\sigma=90$ arcsec.  The light blue areas in this image (where the X-ray flux is low) correspond to regions with high Galactic column density of gas and dust, which attenuate X-rays. {Locations of supernova remnants 3C396, 3C397, and G038.7-01.3 \citep[e.g.,][]{2025JApA...46...14G}, as well as of a HII region Sharpless~74 \citep[][]{1959ApJS....4..257S} are marked on both images.}}
    \label{f:i_survey}
\end{figure*}

~~~~ Since W50 spans approximately $2.2 \times 1.0$ degrees on the sky, corresponding to a physical size of $\sim190 \times 90\,{\rm pc}$ at a distance of 5 kpc\footnote{The most recent parallax measurement for SS~433 in Gaia DR3 corresponds to much larger distance of $8.5^{+2.0}_{-1.4}$ kpc \citep[Gaia~DR3~4293406612283985024,][]{2023A&A...674A...1G}; we advise caution if using this value given the large residuals of the astrometric solution. The previous parallax measurement from {Gaia}~DR2 data gave a much smaller distance of $3.8^{+0.8}_{-1.1}$ kpc \citep[][]{2021MNRAS.502.5455A}.},  {the observing campaign} consisted of nine individual telescope pointings (each covering a circular area of $\sim 1^\circ$ in diameter). These pointings were arranged in a mosaic to give a nearly uniform coverage of the entire nebula and, in addition, the highest sensitivity in the most interesting regions. The observing dates, from October 5 to October 8 and then from October 19 to October 20, 2019, {were chosen} to coincide with the two consecutive orbital eclipses of the central source SS~433 by its normal companion star \citep[according to orbital ephemerids by ][]{2001ApJ...561.1027E}. During eclipses, the apparent X-ray flux of SS~433 goes down by a factor of several \citep[e.g.,][]{1989PASJ...41..491K,2006ApJ...650..338L,2019AstL...45..299M}, minimizing possible adverse effects of contamination and pile-up due to the presence of a very bright object in the field of view, even though the relative suppression of the soft X-ray lines was rather minor \citep{2013ApJ...775...75M}. The typical effective (i.e., equivalent to the nominal eROSITA on-axis performance) exposure time across the map amounts to 30-50 ks, {leading to images a factor of ten deeper} compared to the previously available full-extent maps of W50 by the $ROSAT$ observatory in the 0.5-2 keV band \citep{1996A&A...312..306B} and \textit{Einstein} observatory in the 2-4 keV band \citep{1983ApJ...273..688W}. In addition to that, the eROSITA data have much better spatial ($\sim15"$ on the telescope axis and $\sim 28"$ averaged across the field of view) and spectral ($\sim 65$ eV at 1 keV) resolution \citep{2021A&A...647A...1P}. A comparable quality of the data is provided by multiple \textit{Chandra} and \textit{XMM-Newton} observations, but their coverage of the whole W50 nebula is nonuniform and still incomplete after more than 20 years of observations \citep{2005A&A...431..575B,2022ApJ...935..163S,2024ApJ...975L..28C}. On the other hand, the SRG/eROSITA data are complementary to these observations, as well as to the imaging observations in harder X-rays with \textit{NuSTAR}, which offer a unique view of nonthermally emitting populations. In this paper, we focus on the SRG/eROSITA X-ray data alone and leave combined analyses for future studies. 

We also supplement the PV data with the data obtained in the course of the SRG/eROSITA All-Sky Survey \citep{2021A&A...647A...1P,2021A&A...656A.132S}. Although the survey data are much shallower (the typical exposure time per point is $\sim1000$~s after combining the data of four scans), they provide uniform coverage of the entire area surrounding the nebula. 
For them, the standard data reduction, background estimation, and exposure and vignetting corrections were applied (similarly to mapping diffuse emission from nearby galaxy clusters \citep{2021A&A...651A..41C,2023A&A...670A.156C}, supernova remnants \citep[][]{2021MNRAS.507..971C,2023MNRAS.521.5536K,2024A&A...689A.278K}, and the Galactic diffuse emission \citep{2020Natur.588..227P,2022MNRAS.509.6068K,2024A&A...688A...4C} on even larger scales). 
Here, we also went one step further and combined the PV and all-sky data in a seamless fashion, similarly as it has been done for \textit{Chandra}, \textit{XMM-Newton} and SRG/eROSITA observations of the Perseus cluster \citep{2025arXiv250719987C}. 

A comparison of the $4\times4$ degree fields in radio and X-ray bands is shown in Fig.~\ref{f:i_survey}. Although in the radio image the outer boundary of the W50 nebula appears to be very prominent, the X-ray image is instead dominated by the very bright central object (SS~433) and elongated structures known as extended X-ray jets (EXJs) located well inside the radio boundary of W50 \citep{1980Natur.287..806S,1983ApJ...273..688W,1996A&A...312..306B}. At a distance of 5 kpc, these images span 350 pc {and range from 17 pc to almost 370 pc in terms of distance from the Galactic plane. Given that the scale heights for the cold and {warm} components of the interstellar medium are tens and hundreds of pc, respectively,} one could naturally expect signatures of the environment stratification \citep[e.g.,][]{2011MNRAS.414.2838G}. This is indeed revealed by the asymmetry of the W50 lobes, with the one directed toward the Galactic plane (the western lobe in equatorial coordinates) being more compact than the opposite one (the eastern lobe).

On larger scales, clear variations of the background and foreground diffuse X-ray emission are seen. An inspection of 3D reddening maps \citep[e.g.,][]{2019ApJ...887...93G} showed a clear anticorrelation of the reddening and soft X-ray surface brightness (see Appendix~\ref{s:absorption}), suggesting that absorption of X-rays is the main reason for these variations. Morphologically similar structures appear in the 3D reddening maps at a distance of 2-3 kpc, i.e., in the foreground to W50/SS~433. Therefore, it is plausible that the majority of them are physically unrelated to W50.

\section{Broad-band imaging}
\label{s:broadband}

\begin{figure*}
\centering
\includegraphics[width=0.99\textwidth,angle=00,frame, trim=0.05cm 2.85cm 0.05cm 2.85cm]{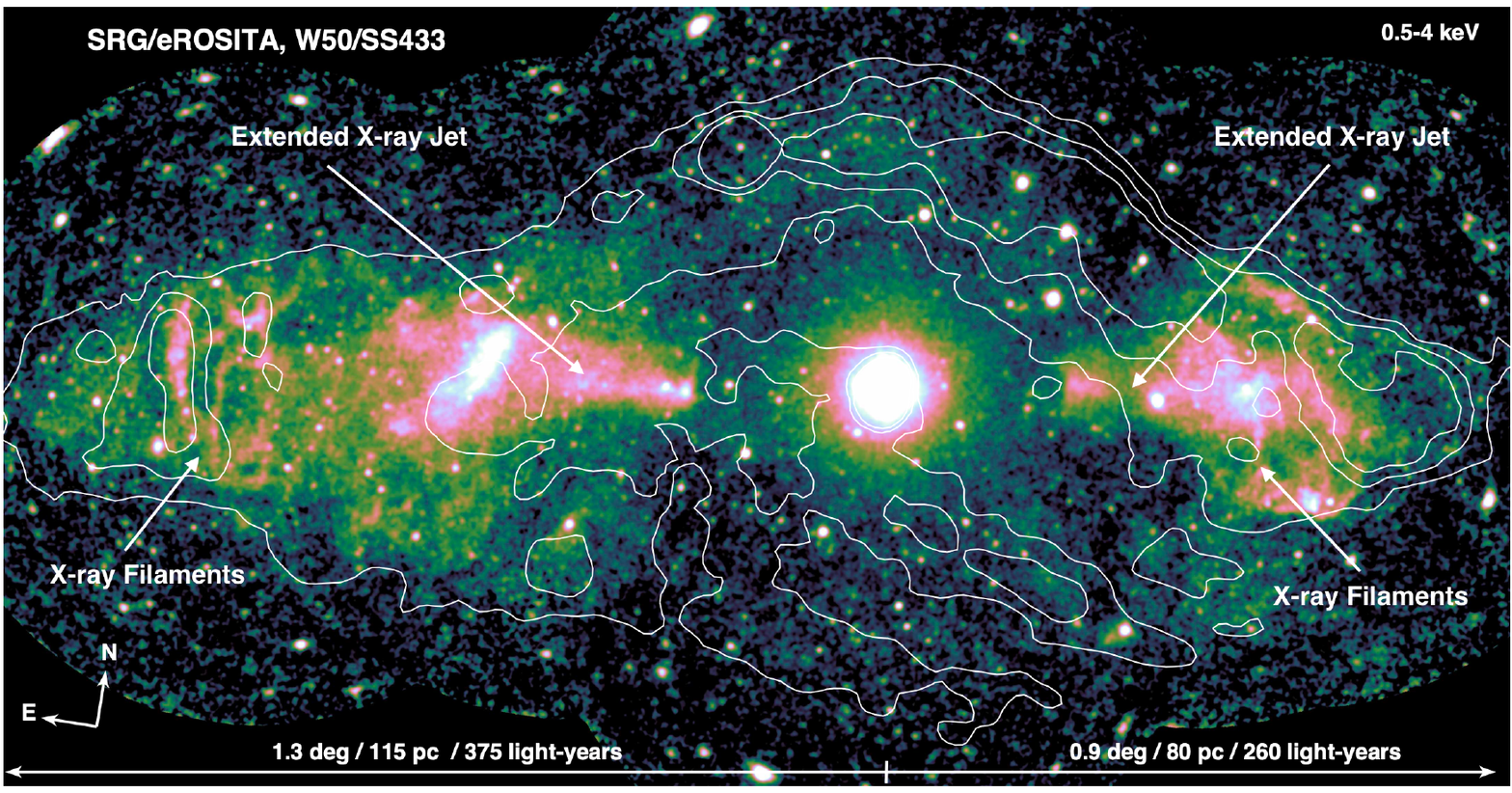}
\caption{Broad-band, high-dynamic-range 0.5-4 keV X-ray image of the giant radio nebula W50 by $SRG$/eROSITA (PV data). The size of the image is 2.2 degrees by 1.1 degrees, corresponding to $\sim200\times100$ pc at a distance of W50, $\sim5$ kpc. The image is rotated so that the longer side of W50 is aligned with the horizontal axis. Color-coding corresponds to the X-ray surface brightness on the square-root scale spanning two orders of magnitude. The bright spot in the center is SS~433. It appears extended due to the wings of the telescope's point spread function around SS~433, which itself is heavily saturated in this image.
The white contours show the surface brightness of the radio emission from the nebula at 4.85~GHz \citep{1996ApJS..103..427G}. With these deeper data (compared to the all-sky survey's data), \revision{a faint diffuse emission is seen between bright extended X-ray jets (EXJs), filamentary structures, and the radio boundary of W50.}}
\label{f:i_broad}
\end{figure*}

We now turn to the deeper PV data. The broad-band (0.5-4 keV) vignetting-corrected X-ray image of the entire W50 nebula is shown in Fig.~\ref{f:i_broad}, with the contours of radio emission at 4.85 GHz \citep{1996ApJS..103..427G} overlaid in white. The color-coding reflects the measured X-ray surface brightness on a square-root scale spanning two orders of magnitude, highlighting the wide dynamic range of the diffuse X-ray emission captured by eROSITA. In addition to SS~433 itself (the extremely bright source in the center) and a multitude of point sources --the majority of which are nearby foreground stars or background active galactic nuclei 
(AGNs) -- the picture shows bright diffuse emission with rich internal structure. 

The X-ray picture can be broadly decomposed into three components. First, there is a low-surface-brightness emission filling the entire projected extent of the nebula. It is brighter in the eastern and western lobes of W50 (left and right, respectively, in Fig.~\ref{f:i_broad}) and significantly dimmer along the northern and southern radio boundaries of W50. Remarkably, the central part of W50 (beyond the region contaminated by the PSF wings from SS~433) appears even fainter in X-rays, suggesting a shell-like geometry of the X-ray-emitting gas. Second, one can notice the presence of very narrow X-ray filaments, particularly inside and close to the boundaries of the eastern and western lobes. Their apparent width is $\gtrsim 1\,{\rm pc}$, while their length is $\sim10\,{\rm pc}$. {Given the angular resolution of eROSITA (better than $\sim 28"$; i.e., $\lesssim 0.7\,{\rm pc}$ at 5 kpc), these filaments are fully resolved in both directions.}  The X-ray filaments form a complicated network with an overall structure very similar to the filaments of the cold and dense gas observed via H$\alpha$+[N II] emission at different locations within the nebula \citep{2007MNRAS.381..308B,2017MNRAS.467.4777F}. The largest X-ray filaments have a flat transverse profile of the surface brightness and are an order of magnitude brighter than the surrounding medium. Such filaments of the dense X-ray-emitting gas might arise as instabilities in the sheared flows, caustics of the strong shock waves, or due to heating of the preexisting cold-gas filaments by hot gas and relativistic particles inside the lobes.
 
The final and perhaps the most visual feature of the image is a pair of nearly symmetrical, sharp-edged 
EXJs directed away from the central source. Their axis is not only aligned with the elongation of the radio nebula and location of the X-ray bright lobes, it also coincides with the projection of the precession axis of SS~433's famous trans-relativistic baryonic jets. The physical nature of these structures, first found by the \textit{Einstein} observatory \citep{1983ApJ...273..688W} and then observed with all major X-ray telescopes \citep{1996A&A...312..306B,1997ApJ...483..868S,2000AdSpR..25..709N,2007A&A...463..611B,1999ApJ...512..784S}, including the most recent polarimteric observations with \textit{IXPE} \citep{2024ApJ...961L..12K}, remains a matter of debate \citep[e.g.,][]{2024A&A...688A...4C}. The $SRG$/eROSITA data clearly show a very rapid onset of emission at the innermost boundary of EXJs on both sides of the nebula {\citep[{also demonstrated recently using \textit{Chandra} data by}][
]{2025ApJ...993L..24T}}.

\section{Spectrally decomposed imaging}
\label{s:specresolved}

\begin{figure*}
\centering
\includegraphics[width=0.99\textwidth,angle=00,bb = 49 295 542 549,clip]{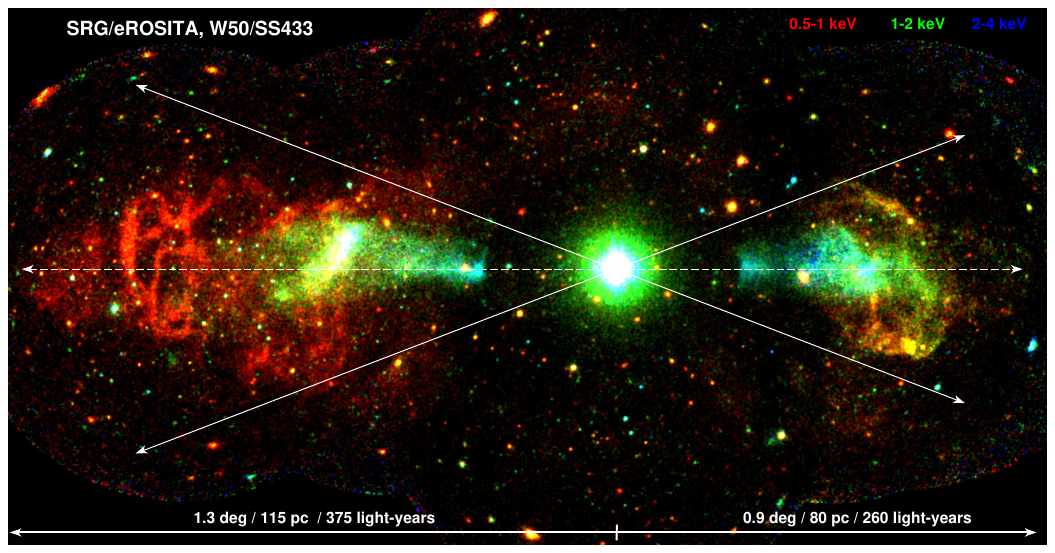}
\caption{Composite X-ray image of the radio nebula W50. The surface brightnesses of the X-ray emission (on a linear scale) in 0.5-1 keV (red), 1-2 keV (green), and 2-4 keV (blue) energy bands are color-coded. The white arrows depict the projection of the precession cone of the SS~433 jets extrapolated to distances of $\sim 100\,{\rm pc}$. Hard and soft X-ray diffuse emission {splits convincingly into two components}: softer filamentary emission (red-yellow) and harder (green-blue) emission of EXJs. On top comes a multitude of nearby (active stars and accreting white dwarfs) and distant (mostly AGNs) compact sources. For distant sources, the absorption by the Milky Way gas suppresses emission below 1 or 2 keV, giving them a blueish color. }
\label{f:rgb}
\end{figure*}

\begin{figure*}
\centering
\includegraphics[width=0.925\textwidth,angle=00,bb = 50 300 540 550,clip]{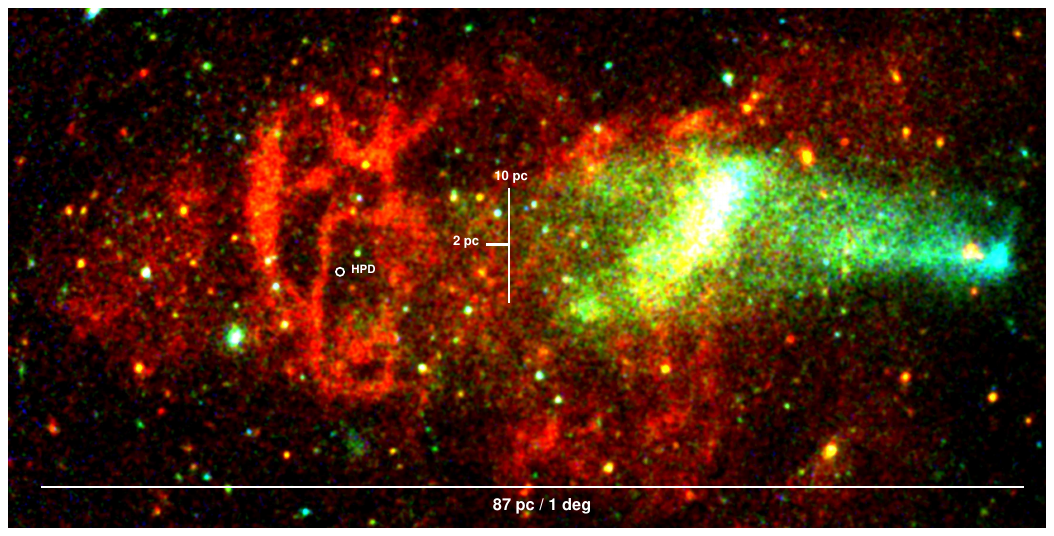}
\includegraphics[width=0.925\textwidth,angle=00,bb = 50 300 540 550,clip]{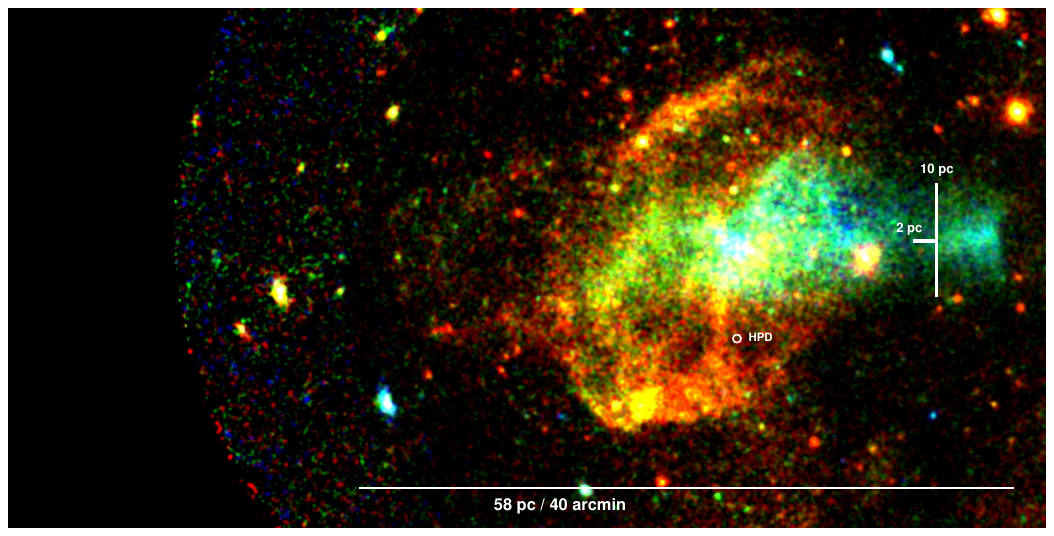}
\caption{Comparison of eastern and western lobes of W50 (see the full image in Fig.~\ref{f:rgb}). The western lobe has been rotated by 180 degrees for the sake of comparison. The white circles show the half-power diameter of the eROSITA PSF {corresponding to the physical scale of 0.7 pc}. Both EXJs emerge at essentially the same distance from the central source with a remarkably sharp edge and {continue 
for $\sim 30$ pc}. On the contrary, the more diffuse 
extensions have different lengths, plausibly associated with much higher ambient density in the western direction.}
\label{f:rgblobes}
\end{figure*}

We took advantage of the unique possibility provided by $SRG$/eROSITA to probe the spectral characteristics of diffuse X-ray emission throughout the full extent of the nebula for the first time.
To produce a spectrally decomposed image,  we divided the full spectral band presented earlier into three subbands: soft (0.5-1 keV), medium (1-2 keV), and hard (2-4 keV) bands, respectively. A composite pseudo-color (red-green-blue)
image showing the surface brightness of the emission (now on a linear scale) in these bands in red, green, and blue, respectively, is presented in Figure \ref{f:rgb}. The chosen spectral decomposition immediately reveals the presence of two distinct components in the observed diffuse emission: large regions in red-orange where emission below 1 keV dominates, and green-cyan structures where prominent emission above 1 keV is present (including the bright central source SS 433). Compact sources also form two groups: the softer (red-yellow) ones are predominantly foreground stars, while harder (blueish) sources are mostly background 
AGNs, which shine through the gas and dust in the disk of the Milky Way. {Zoomed-in images of the lobes (rotated and put side by side to aid comparison) are shown in Fig.~\ref{f:rgblobes} and demonstrate fine structure of the soft diffuse emission, contrasting with relatively smooth and well-confined hard emission.}

\centerline{}

\begin{figure*}
\includegraphics[width=0.99\textwidth,angle=00,bb = 49 295 550 549,clip]{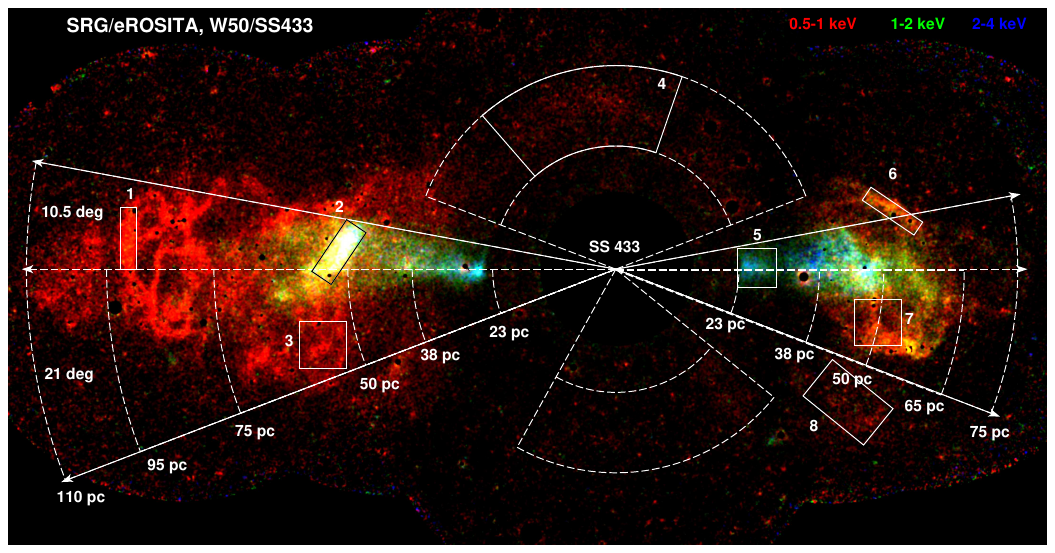}
\caption{"Anatomy'' of diffuse X-ray emission inside W50. The image shows a composite map of the X-ray surface brightness in 0.5-1, 1-2, and 2-4 keV energy bands (similar to Fig.~\ref{f:rgb}) after masking bright foreground and background point sources, and the central source SS~433. Overlaid in white are markers of the distance from the central source (assuming $d_{\rm SS~433}=5$ kpc) and angle with respect to its orbital plane axis, coinciding with the precession axis of its narrow relativistic jets. The box regions, numbered from 1 to 6, are used for spectrum extraction and represent fiducial spectral components composing the image.}
\label{f:rgbdiff}
\end{figure*}

For better characterization of the diffuse emission, we identified bright point sources and masked them along with the central source SS~433. The resulting composite map of the residual diffuse emission is shown in Figure \ref{f:rgbdiff}. The overlaid distance and angle rulers highlight the characteristic dimensions and morphology of the prominent structures. Namely, one can see that the EXJs' start $\sim23$ pc away from the central source and demonstrate remarkable symmetry, with the eastern one starting only a couple of parsecs further from the central source. Both EXJs, characterized by hard X-ray spectra, terminate at around $R\sim60-65$ pc. This symmetry contrasts with the more asymmetric and irregular soft emission of the lobes.

    
%
%

%

{The same conclusion is even better illustrated by the side-by-side comparison of the two lobes in Fig.~\ref{f:rgblobes}.} Indeed, while the eastern lobe extends up to $\sim 110 \,{\rm pc}$, the western one terminates at $\sim 75$ pc. The inclination of the relativistic jet's precession axis to the line of sight is $78^\circ$ (the relativistic jet on the eastern side is on average approaching us), so this difference cannot be explained by geometrical effects alone. However, it broadly agrees with the assumption that the energy flow from the central source interacts with the denser ambient medium in the western part, reflecting the gas density gradient perpendicular to the equatorial plane of the Galactic disk \citep[e.g., ][]{2011MNRAS.414.2838G}. Since the difference between the lobes' dimensions amounts to a sizeable fraction of them, one can infer the exponential scale height of the ambient gas density profile of $\sim100$ pc, which is consistent with the scale height of the Galactic neutral-gas disk. The X-ray emission from the northern and southern boundaries of the nebula appears soft and faint and is noticeably affected by the foreground dust absorption, especially in their western parts (as illustrated in Appendix~\ref{s:absorption}). 

Taking this into account, we conclude that the soft X-ray emission can be traced across the full extent of the nebula, except for the ``inner cavity,'' with a radius of $\sim 23-25$ pc.
The outer boundary of the emission of the northern and southern arcs {(outlined as dashed annuli including Region~4 in Fig.~\ref{f:rgbdiff})} is located at $\sim 38\, {\rm pc}$ from the center, coinciding with the boundary of W50's radio emission. This morphology suggests that the soft X-ray-emitting gas is confined in a thick ($\Delta R\sim R/10$) close-to-spherical shell, driving a strong shock in the surrounding ISM.

The eastern and western lobes are roughly confined inside the projection of the cone with the half-opening angle of $21^\circ$, while the EXJs only encompass the cone with an opening angle that is two times smaller, i.e., $\sim10^\circ$ \revision{(see the characteristic angles indicated in Fig.~\ref{f:rgbdiff}). 
The opening angle of the baryonic jets launched by the central source is much smaller: $\theta_{\rm j}\lesssim2^\circ$ \citep[e.g., ][]{2002ApJ...564..941M}.
Therefore, if the jet direction precesses slowly and the ejected matter moves ballistically, it would only fill thin “walls” of a hollow cone, making a tightly wound spiral with $\beta_{\rm j} P\approx1.1\times10^{17}$~cm $\approx0.036$ pc separation ($\sim1.5''$ at $d_{\rm SS433}=5$~kpc).
Due to a light-travel-time delay between the approaching and receding parts of this spiral, however, the apparent spacing between individual twists should be completely smeared out at distances beyond 1 pc of the central source \citep[e.g., ][]{2017AstL...43..388K}}. From the eROSITA spectrally decomposed pseudo-color (red-green-blue) images, a gentle spectral evolution of the EXJs' emission with distance is seen clearly. EXJ emission is significantly harder at its sharp inner boundary than $\sim 30\,{\rm pc}$ further away from SS~433, where EXJs blend with softer thermal emission that fills the rest of the lobe; this is in agreement with earlier findings of \textit{XMM-Newton} and \textit{NuSTAR} \citep[][]{2007A&A...463..611B,2022ApJ...935..163S}.

\section{X-ray spectra}
\label{s:spectroscopy}

\subsection{Central source} 
\begin{figure}
\includegraphics[width=0.99\columnwidth,trim=1.2cm 9.5cm 1.7cm 3.5cm,clip=true]{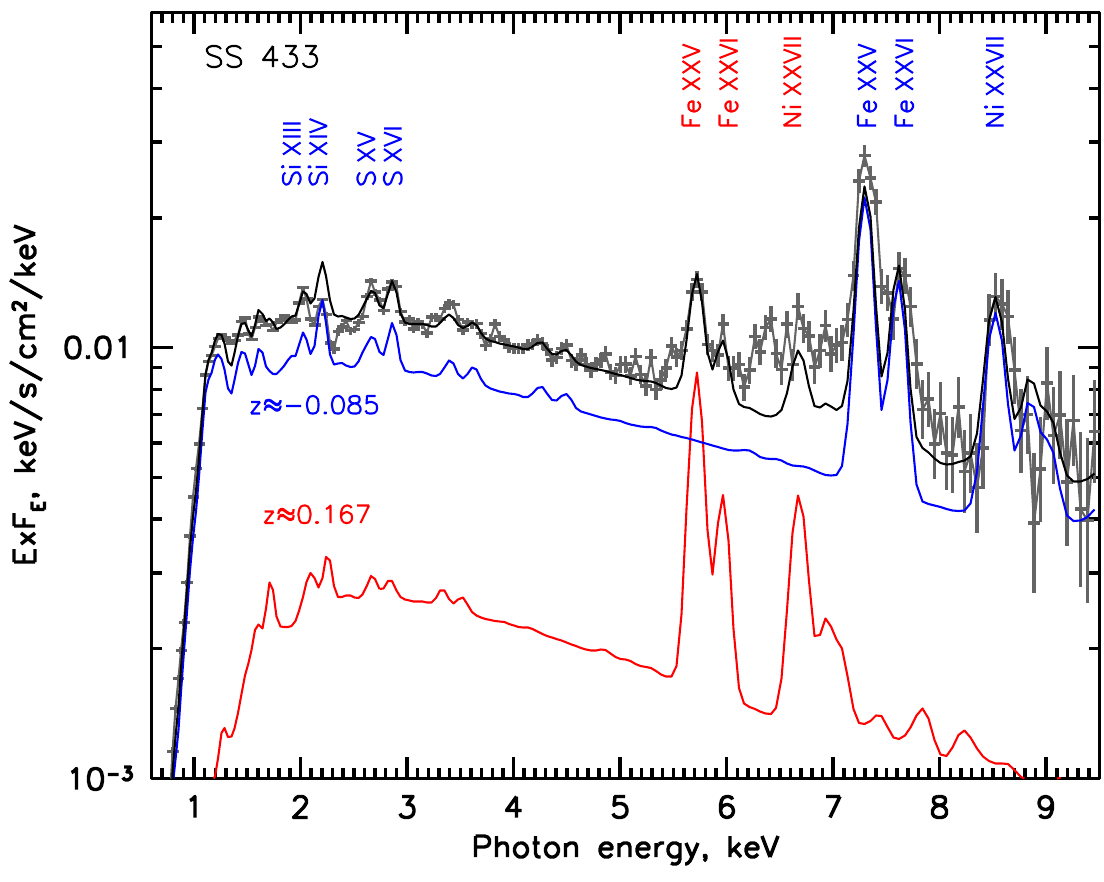}
\caption{Combined spectrum of SS~433 accumulated over PV phase observations.  Both blue-shifted and red-shifted lines of highly ionized (H-like and He-like) heavy elements (in particular, silicon, sulfur, iron, and nickel) are visible and come from the approaching and receding collimated baryonic jets. The model of multi-temperature cooling jets \citep{2016MNRAS.455.1414K} is able to the fit the data very well, although some excess emission is observed between 6 and 7 keV, probably coming from lower ionization species of iron existing in the accretion disk funnel or wind \citep[e.g.,][]{2002ApJ...564..941M,2005A&A...431..575B,2010MNRAS.402..479M,2019AstL...45..299M}.
}
\label{f:ss433_spectra}
\end{figure}

{
The spectral capabilities of eROSITA have been demonstrated using many objects, including SS~433 itself \citep[e.g.,][]{2021A&A...656A.132S}.
Since this source is extremely bright, the observations were intentionally performed during the two consecutive orbital eclipses of the compact object (and the jets) by the companion star, so its soft X-ray flux was dimmer by a factor of several compared to the typical levels \citep[][]{1989PASJ...41..491K,2006ApJ...650..338L,2013ApJ...775...75M}.  The combined X-ray spectrum of SS~433 extracted from an 
{$r=3$ arcmin aperture} (with background estimated from $r=4$ to 5 arcmin annuli; this is, however, completely negligible compared the source) using the data of all pointings in which it falls within the field of view is shown in Fig.~\ref{f:ss433_spectra}.

{As expected, }it is well described by the model of the optically thin multi-temperature emission from collimated flows that adiabatically expand and cool at distances from a few $10^{10}$ to a few $10^{12}$ cm from the central object \citep{1988A&A...196..313B,1991A&A...241..112B,2005A&A...431..575B,1996PASJ...48..619K,2002ApJ...564..941M,2010MNRAS.402..479M,2016MNRAS.455.1414K,2019AstL...45..299M}. {We discuss this further in Appendix~\ref{app:ss433}; the main focus of this study is instead on the much fainter diffuse emission from the surrounding nebula. }

\begin{figure*}[t!]
\centering
\includegraphics[width=0.49\linewidth,bb=30 180 540 670]{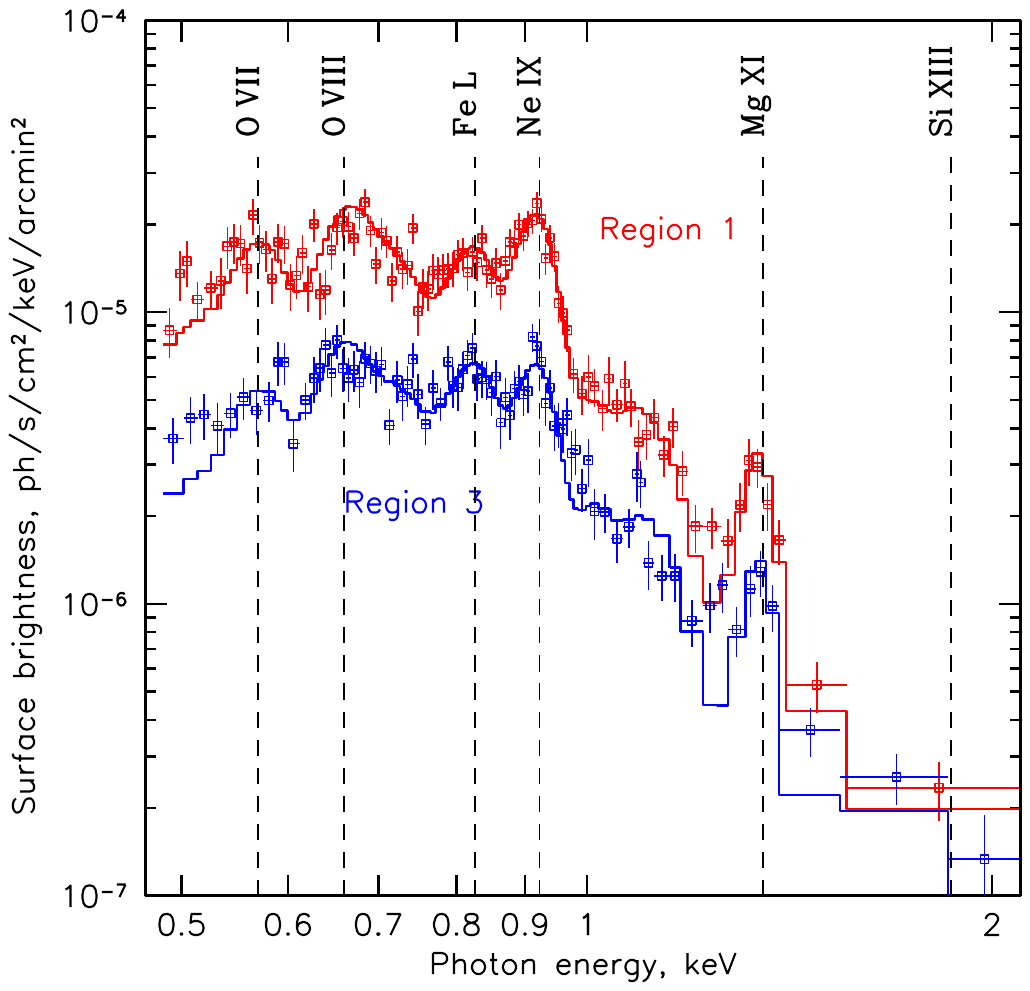}
\includegraphics[width=0.49\linewidth,bb=30 180 540 670]{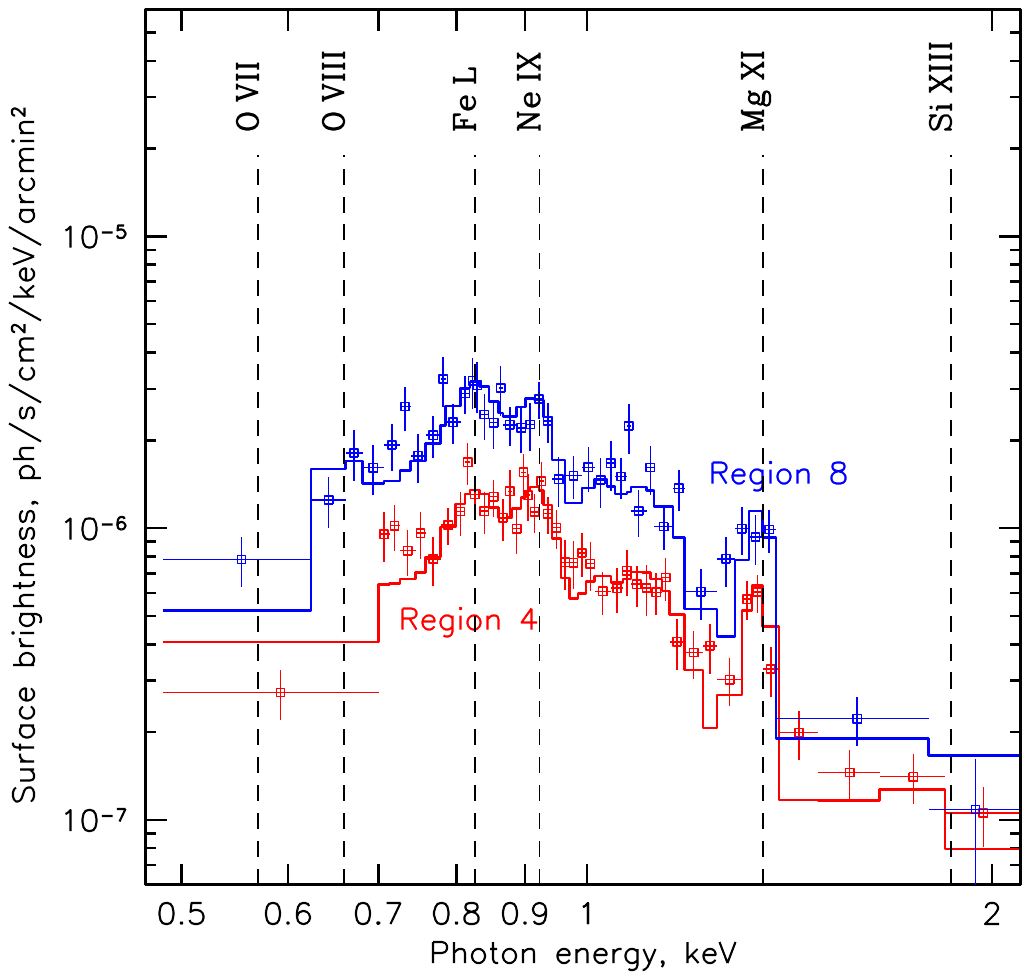}
\includegraphics[width=0.49\linewidth,bb=30 180 540 670]{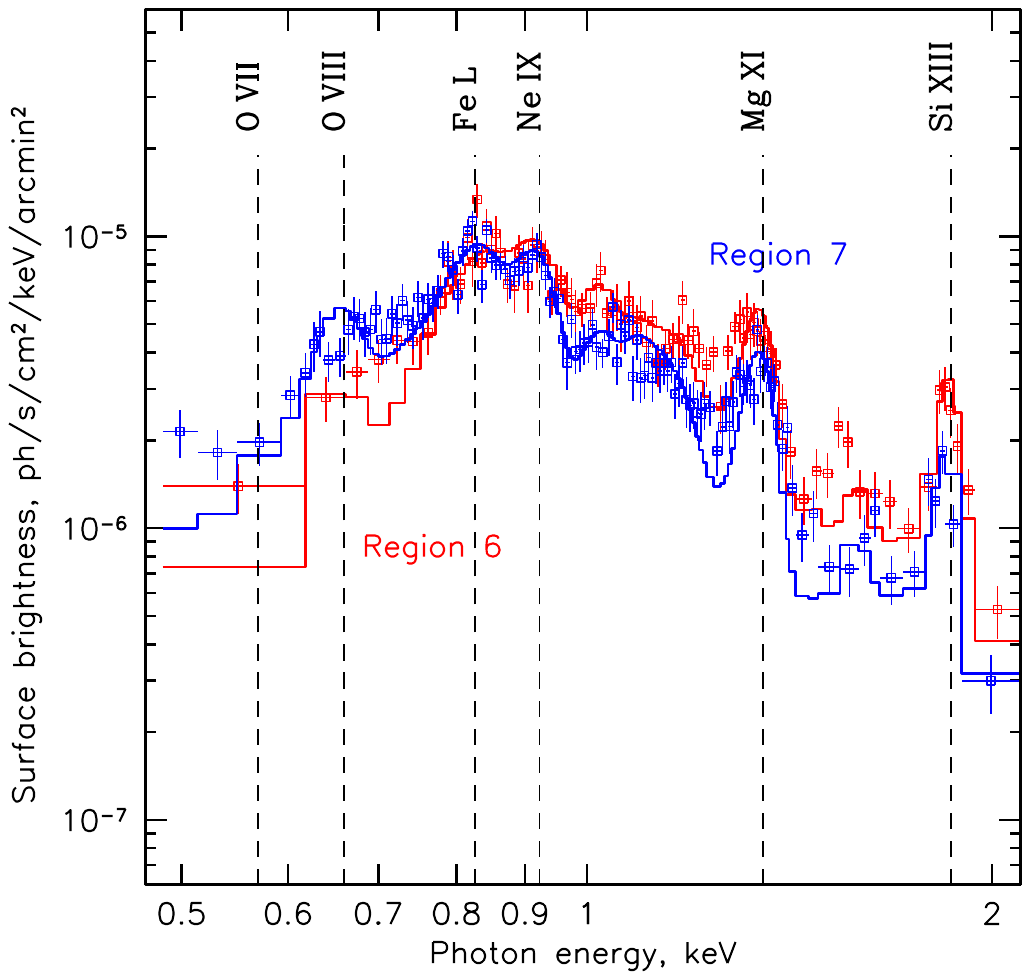}
\includegraphics[width=0.49\linewidth,bb=20 180 540 670]{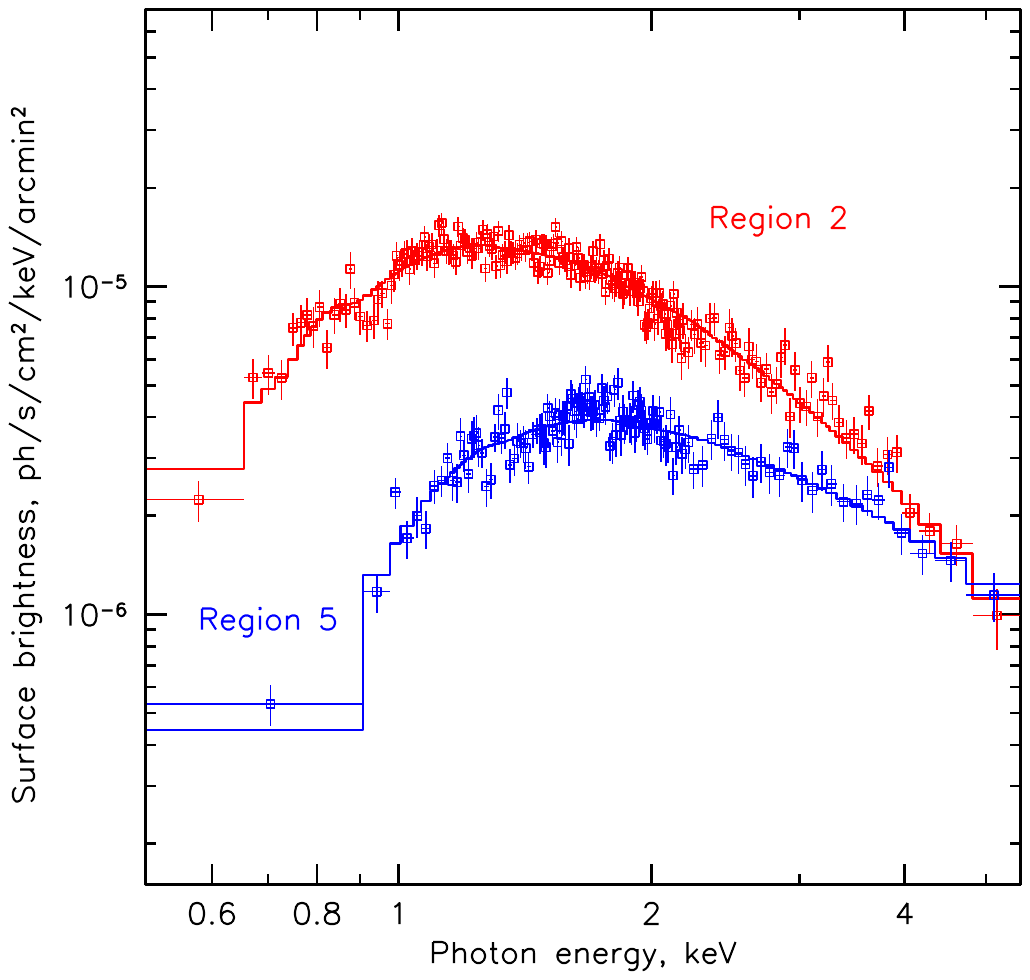}
\caption{Results of spectral analysis for the data from eight representative regions. {\bf Top row.} Left panel shows background-subtracted spectra of the brightest X-ray filament (Region 1 in Figure \ref{f:rgbdiff}, red points with 1$\sigma$ error bars) and the diffuse emission (Region 3, blue points) in the eastern lobe.
{The right panel shows spectra of the northern Arc (Region 4, red points) and of the southern extension (Region 8, blue points).
 {\it Bottom row.} Left panel shows spectra of the diffuse emission in the western lobe (Region 7 in Figure \ref{f:rgbdiff}, blue points) and of the brightest filament in the western lobe (Region 6, red points).} Right panel shows spectra for the termination region of the eastern EXJ (Region 2, red points) and the base of the western jet (Region 5, blue points).
}
\label{f:spectra}
\end{figure*}

\subsection{Diffuse emission} 

The high sensitivity of the presented data allowed us to perform spectroscopy of the diffuse emission in several regions and determine the physical state of the radiating plasma. We selected eight regions, shown as numbered boxes in Figure \ref{f:rgbdiff}, for a detailed spectral analysis of their X-ray emission. For each region, the background (estimated from the adjacent regions) was subtracted to isolate the net signal. We present the parameters of the fit models in Table~\ref{t:spec} and discuss the implications of the corresponding results below.   

\subsubsection{Eastern lobe} 

Emission from the brightest filament in the eastern lobe (Region 1 in Figure \ref{f:rgbdiff}) features a spectrum characteristic for a thermal (but nonequilibrium) optically-thin plasma with a temperature of $\sim 0.3\, {\rm keV}$, as evidenced by the line emission from the highly ionized oxygen, neon, magnesium, silicon, and iron (see the left panel in Figure \ref{f:spectra}). The line emissivity and ratios are consistent with the solar abundance of heavy elements, but they also indicate a nonequilibrium state of the gas ionization. \revision{In particular, the simultaneous presence of spectral lines from ion species correspond to a broad range of gas temperatures (from helium-like oxygen O~VII to numerous iron L-shell ions), likely reflecting an ongoing ionization process of initially cold and rapidly heated ISM gas (although a multi-temperature mixture of the equilibrium spectra is also capable of reproducing similar features). The obtained best-fit, nonequillibrium spectral model is qualitatively consistent with the expectation for a plane-parallel shell of the gas that experienced a passage of the shock with velocity, \revision{$\varv=\sqrt{\frac{16kT}{3\mu m_p}}\sim 500\, {\rm km~s^{-1}} (kT/{\rm 0.3~keV)^{0.5}}$ (from the Rankine-Hugoniot relations for the temperature and density jump properties)}  $\sim 10^4$~yr ago (taking mean molecular weight $\mu=0.6$).}

The inferred hydrogen number density, $n\sim 0.4$~cm$^{-3}$, is a factor of a few higher than the expected gas density of the interstellar gas $\sim300$ pc away from the plane of the Galactic disk. The spectrum of the diffuse emission in the eastern lobe (exemplified by Region 3 in Figure \ref{f:rgbdiff}) also can be well described by thermal nonequilibrium emission of the gas with a somewhat higher temperature of $\sim0.5$ keV and lower hydrogen number density of $\sim 0.1$ cm$^{-3}$ (see the left panel in Figure \ref{f:spectra}). 

\subsubsection{Western lobe} 

The emission from the western lobe appears to be noticeably harder, partially due to the higher interstellar absorption in that direction. However, it also corresponds to the emission of plasma with genuinely higher temperature, $\sim1$ keV, and a similar hydrogen number density, $n\sim0.1$ cm$^{-3}$ (this is a few times higher for the brightest filament), as illustrated by spectra from Regions 6 and 7 in Figure \ref{f:rgbdiff}). The nonequilibrium ionization model (see the right panel in Figure \ref{f:spectra}) results in a comparable estimate of $\sim 10^4$~yr for the ionization timescale.

The total absorption-corrected luminosity of the diffuse soft X-ray emission from the nebula is $\sim10^{35}$ ${\rm erg~s^{-1}}$ in 0.5-4 keV, which is several tens times smaller than the apparent X-ray luminosity of the central object, $L_{\rm X, SS\,433}\sim$few$\times 10^{36}$ ${\rm erg~s^{-1}}$,\footnote{Due to the beaming of the radiation along the thick-accretion-disk axis (expected in the case of highly supercritical accretion), the actual X-ray luminosity of SS 433 might be orders of magnitude larger and namely comparable to the Eddington luminosity and kinetic luminosity of the jets: $L\sim10^{39}$ ${\rm erg~s^{-1}}$ \citep[e.g.,][]{2001IAUS..205..268F,2006MNRAS.370..399B,2007MNRAS.377.1187P,2010MNRAS.402..479M,2016MNRAS.457.3963K,2021MNRAS.506.1045M,2023A&A...669A.149F}.} and much smaller than the supposed kinetic luminosity of the outflows from the supercritical accretion disk: $L_{k}\sim10^{39}$ ${\rm erg~s^{-1}}$. However, the thermal-energy content of the X-ray-emitting gas in the eastern lobe {(with an estimated gas number density of 0.1 cm$^{-3}$, temperature of 0.5 keV, and volume of $\rm V\sim\pi\times 70~pc\times (20~pc)^2\approx88,000~pc^3$, assuming roughly cylindrical geometry)} is $\sim4\times 10^{50}$ erg, requiring a supply of at least $10^{39}$ ${\rm erg~s^{-1}}$ over $10^4$ years. The mass of this gas amounts to $\sim200 M_{\odot}$, implying that the bulk of it cannot be provided by the outflows from SS 433. If the volume filling factor, $f,$ of the emitting medium is low, however, the estimate for the required mass and energy could be decreased by a factor of $1/\sqrt{f}$, but similarly, the density should be increased, leading to the corresponding {decrease} in the time since the shock passage. That means that the X-ray-emitting gas provides an estimate of the average energy input in the lobes at the level of $\sim10^{39}$ ${\rm erg~s^{-1}}$, which is comparable to the kinetic luminosity of the jets and the Eddington luminosity of the system {(for the mass of the compact object below 10$M_{\odot}$)}. 

A similar estimate can be obtained from the western lobe, implying nearly symmetric energy injection along the nebula's axis at a very high rate over the last few tens of thousands of years. {Given that the temperature inferred from the spectral fits with the nonequilbirum emission is driven mostly by characteristics of the ionization state, it is likely a lower limit on the actual temperature of the emitting gas. Hence, a significant portion of the gas energy content might be effectively hidden, while the required energy input should be increased accordingly.}

\subsubsection{Northern and southern boundaries} 

The emission from the nebula boundaries is exemplified by the northern arc region and the southern extension (Regions 4 and 8 in Figure \ref{f:rgbdiff}). The thermal-nonequilibrium model provides a temperature estimate of $\sim0.8$ keV and low hydrogen number density of $\sim 0.1$ cm$^{-3}$ for the northern arc, and this is even lower for the southern extension. The ionization timescale is, however, similar: $\sim 10^{4}$~yr. \revision{The derived parameters of the nonequillibrium ionization model for the northern-arc region (listed in Table~\ref{t:spec}) are in good agreement with the findings of \citet[][]{2024ApJ...975L..28C}, which studied a similar region using the data of dedicated \textit{XMM-Newton} observations guided by earlier mosaic of the whole W50 nebula by ROSAT PSPC \citep[][]{1996A&A...312..306B}. A great advantage of the SRG/eROSITA data (particularly important for the very faint extended emission from these regions) is significantly more robust control over contributions of the instrumental and astrophysical backgrounds. This allows much tighter constraints to be obtained on the key parameters of interest here.}
The observation of rather faint X-ray emission contrasts with the brightness of the radio emission, which in the northern-arc region is one of the brightest across the nebula. The high degree of linear polarization of the radio emission might indicate that the preferential orientation of the interstellar magnetic field in this region favors efficient acceleration of relativistic particles producing synchrotron emission. We note that the central part of the W50 X-ray map suffers from higher photoelectric absorption than the eastern and even western lobes, as illustrated in Fig.~\ref{f:av}. This excess absorption is likely associated with the foreground gas and has no physical connection to SS~433/W50.  

\subsubsection{Extended X-ray jets}

In contrast to this ``thermal'' picture, the X-ray spectra of X-ray emission from the EXJs demonstrate a featureless spectrum that can be described by a power law with a slope ranging from $\Gamma=1.4$ (at their base; e.g., Region 5 in Fig.~\ref{f:rgbdiff}) to $\Gamma=2.2$ (at the termination points; e.g. Region 2 in Fig.~\ref{f:rgbdiff}). \revision{This picture is consistent with the results of focused spectral studies of the western EXJ, including the data above 10 keV from NuSTAR observations \citep[][]{2022ApJ...935..163S}.}
Also, in agreement with previous spectral studies of this emission \citep[e.g.,][]{2007A&A...463..611B}, the thermal bremsstrahlung model also provides an equally good fit with temperatures from $k T=3$ keV (at the termination) to $15$ keV (at the base). The gas emissivity required to produce the X-ray luminosity at level $10^{35}$~${\rm erg~s^{-1}}$ implies hydrogen number densities of $n\sim0.2$ cm$^{-3}$ in these regions (assuming a filling factor of unity). Clearly, if the thermal nature of this emission were true, it would have a drastically higher pressure compared to the ambient, soft X-ray-emitting gas and would expand sideways supersonically, driving a strong shock wave. Alternatively, the emission might result from the population of relativistic particles accelerated within EXJs. The detection of polarization in X-rays \citep{2024ApJ...961L..12K} and TeV emission from regions cospatial with the EXJs \citep{2018Natur.562...82A,2024ApJ...976...30A,2024Sci...383..402H} strongly supports the nonthermal scenario. 

The absence of any noticeable radio, infrared, or optical emission from these structures would then require substantial self-absorption of radiation at these wavelengths (which is unlikely given that the gas number density is likely very low inside them) or a rather peculiar energy distribution of the relativistic particles with a large fraction of them having energies in excess of tens of TeV. 
Another possibility could be scattering of very bright and beamed emission of the central source \citep[e.g.,][]{2013arXiv1306.4486P,2016MNRAS.457.3963K,2019AstL...45..282K}, but the small optical depth with respect to the Thomson scattering, $\tau_{\rm T}=n_{e}\sigma_{T}L\sim 10^{-5} (n_{e}/0.1~{\rm cm^{-3}})(L/50~{\rm pc})$, and the observed spectral gradient along these structures disfavor this possibility.

\section{Discussion}
\label{s:discussion}

The eastern EXJ was well mapped in X-rays with \textit{XMM-Newton}, \textit{Chandra}, and \textit{NuSTAR} \citep[e.g.,][]{2007A&A...463..611B,2022ApJ...935..163S}, revealing its nonthermal nature. Furthermore, recent \textit{IXPE} observation \citep{2024ApJ...961L..12K} revealed polarized X-ray emission from a patch adjacent to the base of the eastern EXJ. These data suggest that X-ray photons are produced by a synchrotron mechanism with the magnetic field predominantly aligned with the EXJ axis. The eROSITA data show that the bases of both EXJs are located at essentially the same distances from SS~433 and have similarly sharp inner boundaries and nonthermal (featureless) spectra across their entire extents.   

These structures have often been considered as a direct consequence of the relativistic jets' propagation inside the nebula before they deposit the bulk of their kinetic energy in the lobes, energizing them and the whole extent of the nebula \citep[e.g.,][]{2014A&A...562A.130P}. However, the direction of the SS 433 jets is known to precess at the mean axis with an amplitude of $\sim21$ degrees, which is a factor of two larger than the observed opening angle of an EXJ. Since this extended X-ray emission comes from distances of approximately {tens of parsec};
i.e., orders of magnitude further away from the central source than the compact radio and X-ray jets, which are traced up to $\sim $ a few $10^{17}$ cm, the actual connection between them is not obvious \citep[e.g.,][]{2017A&A...599A..77P}. It might be governed by recollimation of the jets' flows at a given intermediate distance from the source \citep{1983ApJ...272...48E,1993ApJ...417..170P}, a historical change in the {jets'} direction \citep{2011MNRAS.414.2838G,2015A&A...574A.143M}, or a different regime of baryonic jet propagation through the magnetized interstellar medium (ISM) 
after their gas ``freezes out''  following recombination due to rapid adiabatic expansion, cooling losses, and lack of ambient interactions \citep{2020MNRAS.495L..51C}. 

If we assume that there were no recent drastic changes in the properties of W50/SS~433 (e.g., temporary quenching of the central source), the overall picture of the ``large-scale energy flow'' in W50/SS~433, supported by X-ray data, can be schematically decomposed into the three major components (illustrated in Fig.~\ref{f:summary}) listed below.
\begin{itemize}
    \item Dark flow: this spreads from a compact source up to the base of EXJs covering the range of distances from $\sim 0.1\,{\rm}$ to $\sim 25\,{\rm pc}$ from SS~433. Over this distance range, there are no visible signatures in the X-ray (or any other) band.
    \item Nonthermal flow: this is from the base to the end of EXJs that have featureless nonthermal spectra (presumably synchrotron emission).
 \item Thermal flow: this represents extended diffuse regions between the W50 radio boundaries and EXJs. Its X-ray emission is typical for an ISM that is shock-heated to sub-keV
 temperatures.    
\end{itemize}

Analytical and numerical models primarily aimed at reproducing W50's radio morphology often assume that a quasi-spherical part of the nebula is a remnant of the supernova explosion, while the elongated structures are directly associated with the energy release by the binary system \citep[e.g.,][]{1980ApJ...238..722B,1980MNRAS.192..731Z,1983ApJ...272...48E,1993ApJ...417..170P,2000A&A...362..780V,2008MNRAS.387..839Z,2011MNRAS.414.2838G,2014ApJ...789...79A,2015A&A...574A.143M,2017A&A...599A..77P,2021ApJ...910..149O}.  In the model considered by \cite{2024A&A...688A...4C}, the entire nebula is powered by the central source. There, an anisotropic outflow/wind from the hyper-Eddington accretor is assumed: a combination of a more powerful collimated outflow along the orbital axis and a quasi-isotropic wind in all other directions.  In this model, the darkflow phase corresponds to the free expansion of the wind, and the onset of the EXJs is associated with the termination shock of the isotropic wind, which initiates shocks in the collimated flow where the pressure rises suddenly. Finally, the thermal part is the shock-heated ISM. In this model, the characteristic shape of the W50 nebula (a quasi-spherical part and two extensions) reflects the anisotropy of the velocity and kinetic power of the wind, with no need for the impact of the narrow baryonic jets at all.   

In all these models, the outer shock, which delineates the radio boundary of W50, is not dissimilar to the shocks seen around middle-aged SNRs or stellar wind-blown bubbles \citep[e.g.,][]{1983MNRAS.205..471K,1998AJ....116.1842D}. What makes W50 ``special'' is the presence of EXJs, which are radio-faint but bright in X-rays and VHE gamma-rays. These structures are also plausibly related to efficient particle acceleration at strong shocks, but with a shock configuration that is rather different compared to typical SNR shocks \citep[e.g.][]{2025PhRvD.112f3017B}
. The remarkable symmetry of the EXJs -- which is very clearly demonstrated by the eROSITA image-- points toward the irrelevance of the ambient ISM properties to the energy and momentum propagation, as well as the particle acceleration regime, meaning that a cocoon or a wind-blown cavity likely encompasses these structures in the current epoch. In this regard, the presence of H$\alpha$-emitting filaments in several places across the nebula might help probe the energetic content of the nebula \citep[e.g.,][]{2007MNRAS.381..308B,2010AN....331..412A,2017MNRAS.467.4777F}. 
Interestingly, the location of the brightest H$\alpha$ filaments does not correspond to any prominent features in thermal X-ray emission. This might be the case when the H$\alpha$ emission comes from the outer boundary of the nebula, where the shock wave is already in the radiative mode. Significantly higher gas temperature obtained for the emission from the western lobe might also be indicative of this process, namely that the radiative shock wave runs against the Galactic density gradient and stalls earlier in this direction, evacuating a smaller volume to be filled by the hot X-ray gas. Given the presence of bright silicon lines in the spectra extracted from these regions, an observation of them with the \textit{XRISM} observatory \citep[][]{2025PASJ..tmp...28T} might provide estimates of the directed and turbulent velocities. A large grasp soft X-ray microcalorimeter mission such as \textit{LEM} \citep[][]{2022arXiv221109827K} or \textit{HUBS} \citep[][]{2023SCPMA..6699513B} would also be perfectly suited for the in-depth spectral exploration of this complex system with multiple interacting components.

\begin{figure*}
\includegraphics[width=0.98\textwidth,angle=00,clip=true,trim=3cm 8.9cm 3cm 0cm]{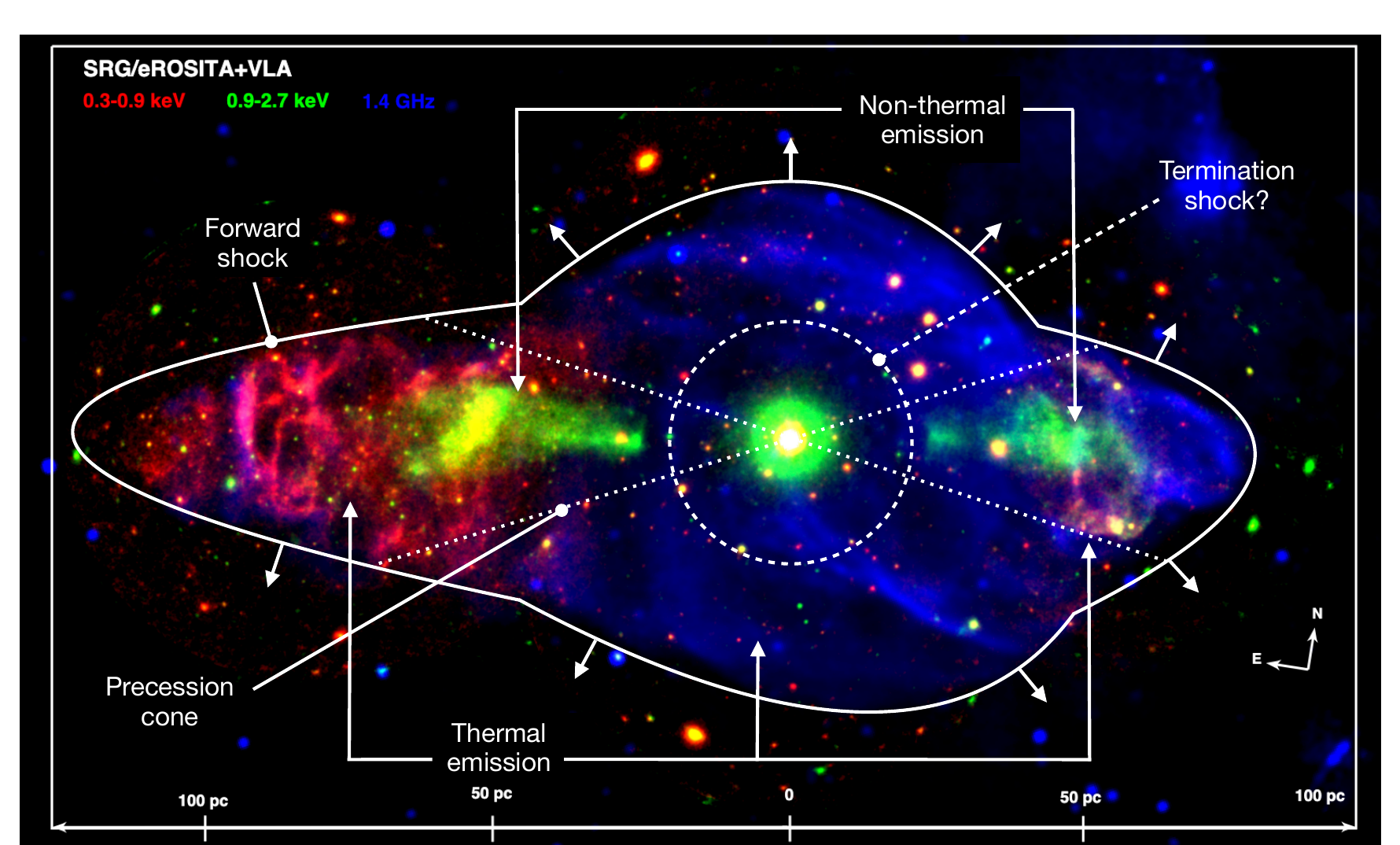}
\caption{A schematic summary picture of the W50 nebula presented on top of the composite X-ray (red and green) and radio (VLA at 1.4 GHz by \citealt{1998AJ....116.1842D}, blue) image. While radio emission most likely arises at the outer shell-like boundary of the nebula, the soft X-ray emission (0.3-0.9 keV) trace shock heated ISM gas behind it which fills almost entire interior of the nebula, and harder X-ray emission (0.9-2.7 keV in this case) is of non-thermal (synchrotron) nature and is produced by ultrarelativistic electrons accelerated at shocks in the axial outflows from the system \citep[e.g.,][]{2024A&A...688A...4C}. The central-most part of the nebula, within $\sim25$~pc from SS~433 (dashed circle) is likely of very low density and could be a wind-blown cavity created by an almost spherically symmetric outflow with close to Eddington kinetic luminosity. The impact of narrow baryonic jets launched by the central source in the current epoch is not clearly visible, given that no indications of interaction are observed along the sky projection of the jet's current precession cone (dotted lines).
}
\label{f:summary}
\end{figure*}

\section{Conclusions}
\label{s:conclusions}
The SRG/eROSITA observations of W50/SS~433 provide, for the first time, the entire X-ray map of the nebula with high spatial and spectral resolution. The data support a physical picture in which the anisotropic energy flow from SS~433 {evolves passing through the following three distinct stages}: 
\begin{itemize}
    \item an invisible dark flow between 0.1 and 25~pc (=anisotropic wind from the binary system)
    \item a nonthermal flow over another $\sim 30\,{\rm pc}$ (=EXJs)
    \item a thermal flow (= shock-heated ISM) that envelopes EXJs
\end{itemize}
The appearance of the nebula is further affected by the large ambient density gradients and a heavy foreground photoelectric absorption that projects almost exactly on the central quasi-spherical part of the W50 nebula.

The thermal part of the W50 X-ray emission can be reasonably well described by the shock-heated plasma that has not yet reached temperature and ionization equilibrium. Such emission is typical for middle-aged or old SNRs. The outer radio boundary of the nebula is also reminiscent of the SNR shocks.

On the contrary,  the EXJs are the most remarkable features of this system on scales of tens of parsec
. Their sharp inner edges plausibly correspond to extreme shocks that accelerate particles and power the X-ray (synchrotron) emission, and, at energy levels that are ten orders of magnitude higher, the TeV
emission. The W50/SS~433 system clearly illustrates the important role the hyper-Eddington accretors might play in the energetics of the ISM in galaxies at different redshifts and the production of ultra-high-energy particles.

\section*{Data availability}
{X-ray data analyzed in this article were used with the permission of the Russian SRG/eROSITA consortium. The reduced broad-band image (Fig.~2) is available at the CDS via anonymous ftp to cdsarc.cds.unistra.fr (130.79.128.5). 
}

\begin{acknowledgements}
    
This work is partly based on observations with the eROSITA telescope onboard \textit{SRG} space observatory. The \textit{SRG} observatory was built by Roskosmos in the interests of the Russian Academy of Sciences, represented by its Space Research Institute (IKI) in the framework of the Russian Federal Space Program, with the participation of the Deutsches Zentrum für Luft- und Raumfahrt (DLR). The eROSITA X-ray telescope was built by a consortium of German Institutes led by MPE, and supported by DLR. The \textit{SRG} spacecraft was designed, built, launched, and operated by the Lavochkin Association and its subcontractors. The science data are downlinked via the Deep Space Network Antennae in Bear Lakes, Ussurijsk, and Baikonur, funded by Roskosmos. 

The development and construction of the eROSITA X-ray instrument was led by MPE, with contributions from the Dr. Karl Remeis Observatory Bamberg $\&$ ECAP (FAU Erlangen-Nuernberg), the University of Hamburg Observatory, the Leibniz Institute for Astrophysics Potsdam (AIP), and the Institute for Astronomy and Astrophysics of the University of Tübingen, with the support of DLR and the Max Planck Society. The Argelander Institute for Astronomy of the University of Bonn and the Ludwig Maximilians Universität München also participated in the science preparation for eROSITA. The eROSITA data were processed using the eSASS/NRTA software system developed by the German eROSITA consortium and analyzed using proprietary data reduction software developed by the Russian eROSITA Consortium.

IK acknowledges support by the COMPLEX project from the European Research Council (ERC) under the European Union’s Horizon 2020 research and innovation program grant agreement ERC-2019-AdG 882679.

\revision{Some of the maps were built using \texttt{cubehelix} colorscheme developed by David Green \citep[][]{2011BASI...39..289G}. }

\end{acknowledgements}

\bibliographystyle{aa}
\bibliography{current} 

\begin{appendix}

\section{Spectrum of the central source} 
\label{app:ss433}

As one can see in Fig.~\ref{f:ss433_spectra}, both blue-shifted and red-shifted lines of highly ionized (H-like and He-like) heavy elements (in particular, silicon, sulfur, iron, and nickel) moving at 0.26 speed of light can be easily discriminated and identified. They arise from the approaching (with apparent redshift $z_{\rm app}=$-0.085) and the receding jet (with apparent redshift $z_{\rm rec}=$0.167), respectively. In agreement with the findings of numerous previous observations \citep[see ][ for a recent summary]{2019AstL...45..299M}, a factor of 10 overabundance of nickel (with respect to iron) in the gas of the jets is required to describe the spectrum well (most clearly indicated by the ratio of the Ni XXVII and Fe XXV lines). Spectral resolution of eROSITA is not sufficient to resolve the width of the lines, which is known to be dominated by the ballistic motion of gas within a cone of with opening angle $\theta_{\rm j}\sim1\deg$, as measured by high resolution spectroscopy with \textit{Chandra} gratings \citep[e.g.][]{2002ApJ...564..941M}. Some residual emission can also be observed in the iron K complex region between 6 and 7 keV, which could come from the reflection of central source emission on the walls of the optically thick accretion disk funnel \citep[][]{2010MNRAS.402..479M} or in the optically thin accretion disk wind \citep[][]{2018AstL...44..390M,2019AstL...45..299M}. Comparison of the spectra taken in the two epochs of observations separated by one orbital period shows very good consistency between them\footnote{{Since one orbital period separation is small compared to the precession period and the observations were close to an extreme point of the precession curve, the change in the Doppler shifts of the jets is small.}}, with the biggest difference observed in the 6-7 keV band. This might be related to the change in the visible geometry of this region or to the physical conditions within it \citep[see a discussion in] []{2019AstL...45..299M}.
High-resolution spectroscopy of this emission with the \textit{XRISM} observatory \citep[][]{2025PASJ..tmp...28T} will open new possibilities for characterizing this emission and potentially give access to more intricate diagnostics based on velocity structure and resonant scattering in the jets \citep[][]{2012AstL...38..443K}, as well as possible signatures of multiphase cooling and ionization state dynamics in supersonically expanding jets prone to thermal instability and fragmentation \citep[][]{1988A&A...196..313B,2000A&A...363..640B,2016MNRAS.455.1414K}.  The SS433's spectrum presented here illustrates the spectroscopic capabilities of eROSITA and shows that, for sufficiently bright sources, reliable continuum and line characterization can be achieved up to 10 keV. 
\section{Parameters of the spectral models}
\label{s:spectralmodels}

 Here we list parameters of the best-fit models used to describe spectra shown in Fig.~\ref{f:spectra} and discussed in Sec.~\ref{s:spectroscopy}. The spectral analysis was performed using standard routines of the \texttt{XSPEC} package \citep{1996ASPC..101...17A}. Namely, the absorbed \citep[using \texttt{tbabs} model by ][]{2000ApJ...542..914W} non-equilibrium plasma emission model \citep[\texttt{nei} model by ][]{2001ApJ...548..820B} was used to describe thermal emission, while similarly absorbed thermal bremsstrahlung model (\texttt{tbabs*brems} in \texttt{XSPEC} notation) was used for a phenomenological description of the non-thermal emission from the EXJs.  Due to the complexity of the system, the X-ray emission from all these regions is likely a mixture of several components; hence, one needs to take these derived parameters with caution and probably as approximate guidelines for the physical conditions within them. For the background estimation, adjacent regions of significant size were used, and the signal inside them was subtracted from the corresponding source region. For all regions except for the relatively faint Regions 4 and 8, the relative contribution of the background emission in the source regions was rather small. 

\begin{table}
\caption{Parameters (with 1$\sigma$ uncertainties) of the best fit models for the spectra extracted from regions shown in Fig.~\ref{f:rgbdiff}.}
\vspace{0.2cm}
\begin{center}
\begin{tabular}{lccccc}
\hline
\hline \\
\vspace{0.2cm}
{\#} & $N_{\rm H}$ & {$kT$} & {$\tau$} & Norm\\ 
\vspace{0.2cm}
& $10^{22}\,\rm cm^{-2}$ & keV & ${\rm 10^{11}~s~cm^{-3}}$& $10^{-5}$   \\
\hline \\
\multicolumn{5}{c}{Thermal regions (\texttt{tbabs*vnei})} \\
\hline \\
\vspace{0.2cm}
1& 0.67$\pm$0.02 &0.29$\pm$0.02 &0.26$\pm$0.05& 12.33$\pm$3.63\\ 
\vspace{0.2cm}
3& 0.56$\pm$0.02& 0.56$\pm$0.07& 0.11$\pm$0.01& 0.80$\pm$0.19\\ 
\vspace{0.2cm}
4& 0.79$\pm$0.05& 0.95$\pm$0.25& 0.09$\pm$0.01& 0.21$\pm$0.07\\ 
\vspace{0.2cm}
6&0.85$\pm$0.03& 0.76$\pm$0.06 &0.64$\pm$0.12 &2.01$\pm$0.28\\ 
\vspace{0.2cm}
7& 0.72$\pm$0.02& 1.30$\pm$0.18& 0.08$\pm$0.01& 0.88$\pm$0.13\\ 
\vspace{0.4cm}
8& 0.75$\pm$0.05& 0.78$\pm$0.22& 0.13$\pm$0.05 &0.42$\pm$0.18\\ 
\multicolumn{5}{c}{Non-thermal regions (\texttt{tbabs*brems})} \\
\hline \\
2& 0.51$\pm$0.01& 3.05$\pm$0.15&-& 1.62$\pm$0.04\\
\vspace{0.2cm}
5& 0.84$\pm$0.03& 14.5$\pm$3.9&-&0.64$\pm$0.01\\
\hline
\end{tabular}
\tablefoot{{The thermal non-equilibrium model used for Regions 1,3,4,6-8 is \texttt{tbabs*nei} with Solar abundance of heavy elements having the absorbing column density $N_{\rm H}$, gas temperature $kT$, ionization timescale $\tau$ and the standard \texttt{XSPEC} emission measure ${\rm Norm=10^{-14}n_{e}n_{H} V/({4\pi d_{SS~433}^2})}$ normalization (per arcmin$^{-2}$) as free parameters. The value of ${\rm Norm}$ converts into the hydrogen number density as $n_{\rm H}={\rm 0.2~cm^{-3}(Norm/10^{-5}})^{0.5}(dl/{\rm pc})^{-0.5}$, where ${dl}$ is the line-of-sight extent of the emitting region (assuming filling factor $f$ equal to unity and $n_{\rm e}=1.21 n_{\rm H}$ for the fully ionized cosmic plasma). For the non-thermal emission model used for Regions 2 and 5 we list parameters of the \texttt{tbabs*brems} fits, where $kT$ is the effective temperature of the bremsstrahlung spectrum. For the normalization, we divided the standard XSPEC normalization of the \texttt{brems} model by the factor 3.28(=$10^{-14}/3.05\times10^{-15}$) to match the normalization definition of the \texttt{nei} model used for the thermal regions. Correspondingly, the same conversion to the number density of the emitting gas can be used in this case as well, demonstrating that would the bremsstrahlung model were true, the required gas density would be comparable to the density of the ambient medium.}}
\end{center}
\label{t:spec}
\end{table}

\section{Impact of foreground absorption}
\label{s:absorption}

{
In this section, we demonstrate that some of the large-scale variations in the diffuse
X-ray emission from W50 is plausibly caused by foreground absorption. To do so, we used the 3D dust maps from Bayestar19 \cite[see][for detailed description]{2019ApJ...887...93G}. We are looking for the intervening absorption between W50 and us, i.e., physically unrelated layers of the absorbing material. Given the adopted distance to SS~433 of 5~kpc, we integrated extinction (${\rm A_V}$) only up to a distance of 3~kpc from the Sun. The resulting extinction map is superposed on the X-ray image in Fig.~\ref{f:av}. A prominent wide dust lane is seen in this image that crosses the central part of the W50 nebula diagonally from NE to SW. Typical values of the extinction in this region are $\sim 5-10$, which translate into equivalent hydrogen column densities $\sim (1-2) \times 10^{22}\,{\rm cm^{-2}}$. Therefore, soft X-ray emission from the central part of W50 can be heavily absorbed (several orders of magnitude at 1~keV). While some geometric correspondences between the X-ray and extinction maps look tantalizing, we conclude that most probably this is a by-chance projection. Therefore, the lack of observed soft X-ray emission from the quasi-circular part of the nebula does not place significant constraints on the properties of W50. The soft X-ray emission might be there, but we can not see it.    

}
\begin{figure}
\includegraphics[width=0.48\textwidth,angle=00,bb=50 150 540 650,clip,frame]{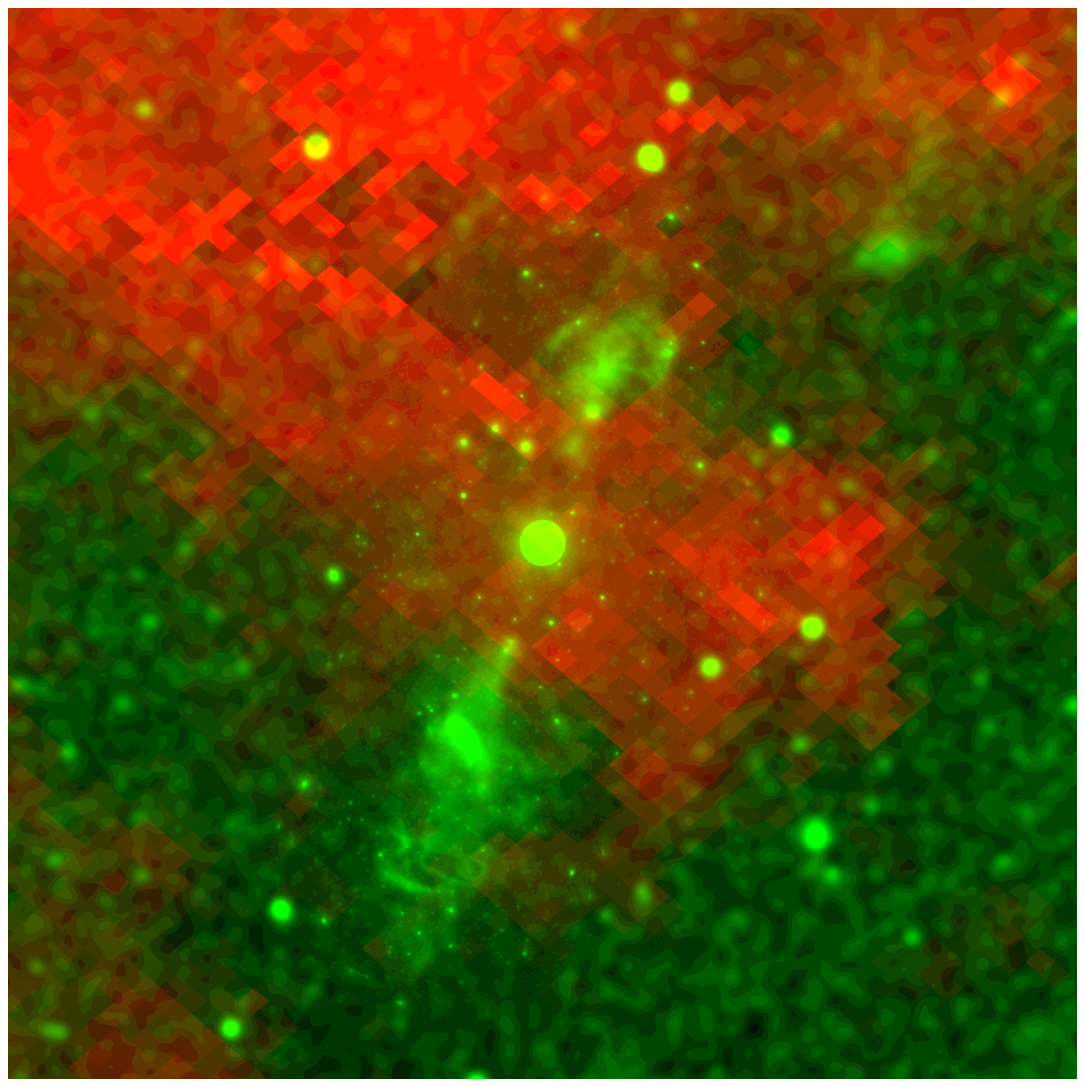}
\caption{X-ray map of W50 in Galactic coordinates with the extinction map (in red) superposed. The extinction is estimated from the Bayestar19 3D maps \cite{2019ApJ...887...93G} up to a distance of 3~kpc from the Sun, i.e., conservatively in the foreground to W50. The extinction (${\rm A_V}$) values range between $\sim 5$ and $\sim 10$ in regions overlapping with the central quasi-spherical part of W50. This apparently accidental projection of the absorbing dust/gas onto W50 leads to significant suppression of the soft X-ray flux in the affected areas.}
\label{f:av}
\end{figure}

\end{appendix}


\label{lastpage}
\end{document}